%% file: main.tex
\documentclass[twocolumn,superscriptaddress,showpacs,nofootinbib,preprintnumbers,secnumarabic,amssymb, nobibnotes, aps, prd]{revtex4-2}
\usepackage[utf8]{inputenc}
\usepackage{graphicx}
\usepackage{latexsym,amsmath,amssymb,amsthm,lmodern,float,url}
\usepackage{natbib}
\usepackage{color}
\usepackage{microtype}
\usepackage{import}
\usepackage{bbold}
\usepackage[plain]{fancyref}
\usepackage{varioref}
\usepackage{slashed}
\usepackage{multirow}
\usepackage{tikz}
\usepackage{scrextend}
\usepackage{braket}
\usepackage{units}
\usetikzlibrary{shapes}
\usetikzlibrary{positioning}
\usepackage[normalem]{ulem}

\newcommand{\fig}[1]{Fig.~\ref{fig:#1}}

\newcommand{\eq}[1]{Eq.~(\ref{eq:#1})}
\newcommand{\Sec}[1]{Sec.~(\ref{sec:#1})}

\usepackage[colorlinks=true,backref=false, linktocpage=true,
citecolor=blue,urlcolor=blue,linkcolor=blue,pdfpagemode=UseOutlines]{hyperref}
\hypersetup{%
  bookmarksnumbered=true,
  pdftitle = {},
  pdfsubject = {},
  pdfauthor = {},
  pdfkeywords = {}
}

\newcommand{\btt}{\mathbb{BT}}

\DeclareMathOperator{\Tr}{Tr}

\def\Tr{\mathrm{Tr}\,}

\newcommand{\be}{\begin{equation}}
\newcommand{\ee}{\end{equation}}
\newcommand{\bea}{\begin{eqnarray}}
\newcommand{\eea}{\end{eqnarray}}
\newcommand{\id}{\mathbb{I}}

\usepackage{amsmath}
\usepackage{amssymb}
\usepackage{mathrsfs}
\usepackage{multirow}
\usepackage{tabularx}
\usepackage{placeins}
\usepackage{graphicx}

\begin{document}
\preprint{\hspace{0.15cm}USTC-ICTS/PCFT-24-47}
\title{Quantum Frontiers in High Energy Physics}
\author{Yaquan Fang}
\affiliation{Institute of High Energy Physics, Chinese Academy of Sciences, Beijing 100049, People’s Republic
of China}
\affiliation{University of Chinese Academy of Sciences, Beijing 100049, People’s Republic of China}
\author{Christina Gao}
\email{gaoy3@sustech.edu.cn}
\affiliation{Department of Physics, Southern University of Science and Technology, Shenzhen, Guangdong 518055, China}
\author{Ying-Ying Li}
\email{liyingying@ihep.ac.cn}
\affiliation{Institute of High Energy Physics, Chinese Academy of Sciences, Beijing 100049, People’s Republic
of China}
\affiliation{Interdisciplinary Center for Theoretical Study, University of Science and Technology of China, Hefei, Anhui 230026, China}
\affiliation{Peng Huanwu Center for Fundamental Theory, Hefei, Anhui 230026, China}
\author{Jing Shu}
\affiliation{School of Physics and State Key Laboratory of Nuclear Physics and Technology, 
	Peking University, Beijing 100871, China}
\affiliation{Center for High Energy Physics, Peking University, Beijing 100871, China}
\affiliation{Beijing Laser Acceleration Innovation Center, Huairou, Beijing, 101400, China}
\author{Yusheng Wu}
\affiliation{Department of Modern Physics and State Key
Laboratory of Particle Detection and Electronics,
University of Science and Technology of China, Hefei 230026, China}
\author{Hongxi Xing}
\affiliation{Key Laboratory of Atomic and Subatomic Structure and Quantum Control (MOE),
Guangdong Basic Research Center of Excellence for Structure and Fundamental Interactions of Matter,
Institute of Quantum Matter, South China Normal University, Guangzhou 510006, China}
\affiliation{Guangdong-Hong Kong Joint Laboratory of Quantum Matter,
Guangdong Provincial Key Laboratory of Nuclear Science, Southern Nuclear Science Computing Center,
South China Normal University, Guangzhou 510006, China}
\affiliation{Southern Center for Nuclear-Science Theory (SCNT),
Institute of Modern Physics, Chinese Academy of Sciences, Huizhou 516000, China}
\author{Bin Xu}
\email{binxu@pku.edu.cn}
\affiliation{School of Physics and State Key Laboratory of Nuclear Physics and Technology, 
	Peking University, Beijing 100871, China}
\author{Lailin Xu}
\affiliation{Department of Modern Physics and State Key
Laboratory of Particle Detection and Electronics,
University of Science and Technology of China, Hefei 230026, China}
\author{Chen Zhou}
\affiliation{School of Physics and State Key Laboratory of Nuclear Physics and Technology, Peking University, Beijing 100871, China}
\date{\today}

\begin{abstract}
Numerous challenges persist in High Energy Physics (HEP), the addressing of which requires advancements in detection technology, computational methods, data analysis frameworks, and phenomenological designs. We provide a concise yet comprehensive overview of recent progress across these areas, in line with advances in quantum technology. We will discuss the potential of quantum devices in detecting subtle effects indicative of new physics beyond the Standard Model, the transformative role of quantum algorithms and large-scale quantum computers in studying real-time non-perturbative dynamics in the early universe and at colliders, as well as in analyzing complex HEP data. Additionally, we emphasize the importance of integrating quantum properties into HEP experiments to test quantum mechanics at unprecedented high-energy scales and search for hints of new physics.  
Looking ahead, the continued integration of resources to fully harness these evolving technologies will enhance our efforts to deepen our understanding of the fundamental laws of nature. 
\end{abstract}

\maketitle

\section{Introduction}

High Energy physics (HEP), commonly referred to as particle physics, is focused on the study of fundamental particles and the principles governing their interactions. The history of HEP is punctuated by significant milestones. Notably, the discovery of the Higgs boson in 2012 \cite{ATLAS:2012yve, CMS:2012qbp} established the Standard Model (SM) as the most successful low-energy effective theory to encapsulate our current understanding of the particles and three of the fundamental forces in its domain. However, many challenges remain unsolved, including the nature of dark matter, the charge conjugation (C) and parity (P) properties of the quantum chromodynamics (QCD) sector, the origin of matter-antimatter asymmetry, the behavior of matter under extreme conditions such as those in neutron stars, and the dynamics of QCD in the strong coupling regime, among others. Continued research in these areas could lead to groundbreaking discoveries and revolutionize our understanding of the fundamental laws of nature.
To effectively tackle these challenges, advancements in detection technology, computational methods, and data analysis frameworks, along with innovative phenomenological studies, need to be pursued, requiring efforts across various fields \cite{p5report}.
\begin{figure*}[t]
\centering\includegraphics[width=0.98\linewidth]{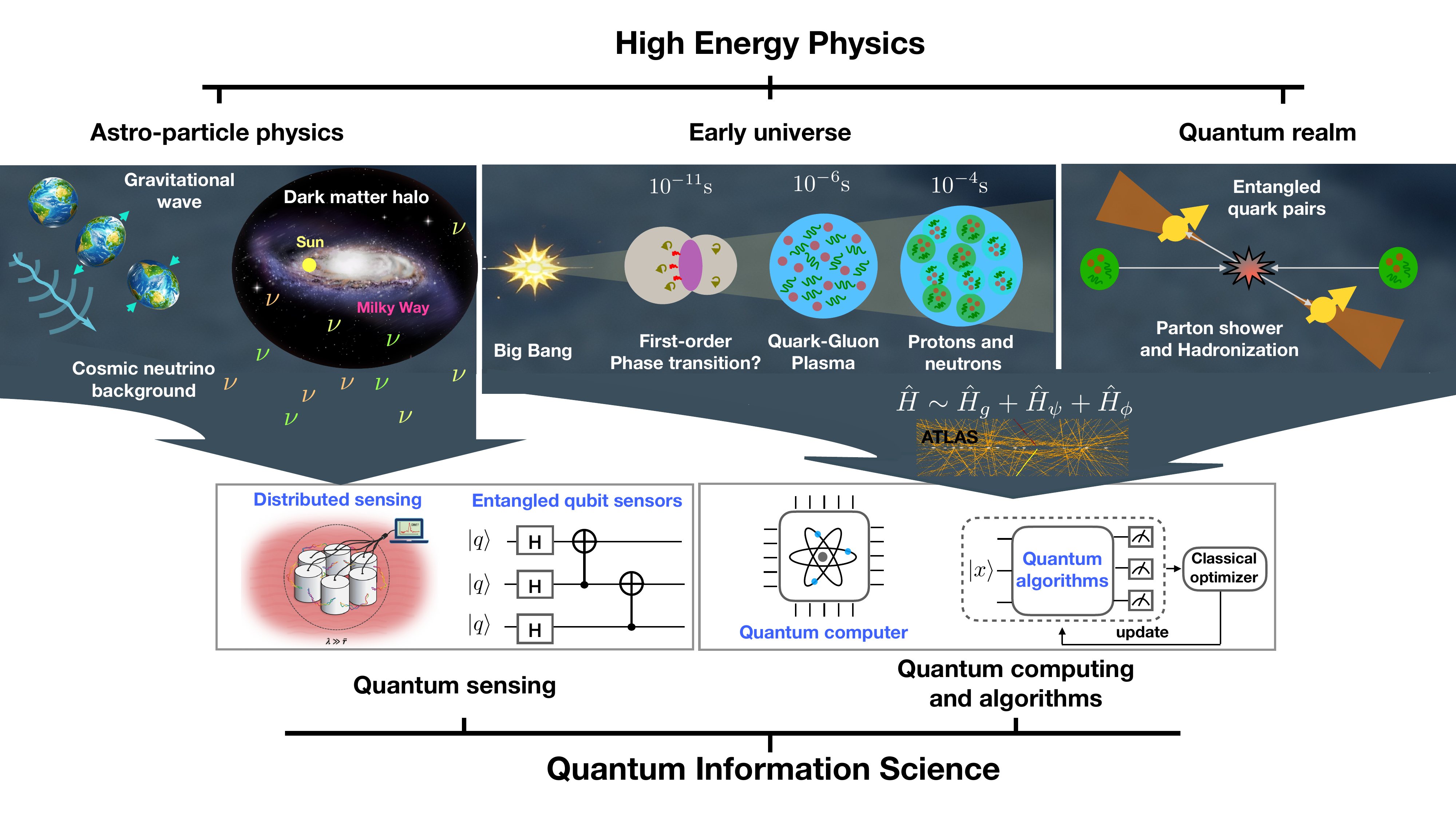}
    \caption{The different areas in HEP and the quantum frontiers that can facilitate their progresses.}
    \label{fig:outline}
\end{figure*}

On the other hand, quantum information science (QIS) has made significant strides. Techniques for controlling single atoms, electrons, and photons have led to innovations in quantum sensors with high sensitivity to weak forces \cite{Chou:2023hcc, Bass-Doser,PRLEssay}. Advancements in techniques for manipulating groups of these particles hold great promise for the construction of scalable quantum computing platforms, along with the development of new quantum algorithms that demonstrate potential speedups over classical algorithms \cite{PRXQuantum.2.017001}. These achievements have not only taken us into uncharted territory but have also transformed our approach to scientific inquiry in HEP.
In this review, we aim to provide a concise yet comprehensive overview of the areas in HEP that are advanced or could be advanced by the developments of QIS, as outlined in \fig{outline}. We will explore how quantum devices, with their unprecedented precision, could detect subtle effects of potential new physics beyond the Standard Model (BSM). Additionally, we will discuss the impact of quantum algorithms and the emergence of future large-scale quantum computers on our theoretical studies of non-perturbative dynamics by solving complex problems in polynomial time, as well as their potential to help uncover new observations by handling the vast and intricate data from HEP experiments. Furthermore, we highlight the importance of incorporating quantum properties into HEP experiments, as this could immediately test our understanding of quantum mechanics at unprecedented high-energy scales and potentially open new avenues to look for hints of BSM from a quantum mechanical perspective.

This lays out the structure of the current review. We dive into recent developments in testing fundamental principles and searching for new physics with quantum sensing and its increased sensitivity in \Sec{sensing}. Then in \Sec{simulation}, we
discuss recent explorations in quantum computing as new paradigms to study dynamics in the non-perturbative regime, including the developments of general framework for QCD calculations and low-dimensional benchmark studies. In \Sec{qml}, we provide a brief summary of the growing efforts to assess the applicability of quantum machine learning in analyzing HEP data and to explore its potential advantages in enhancing the efficiency of uncovering new observations. \Sec{pheno} aims to bring recent results in testing quantum properties at HEP experiments into the spotlight, as well as summarize recent new developments in phenomenological studies. We will then discuss perspectives for future directions to explore and conclude in \Sec{conclusion}.

\input{sections/quantum-sensing}
\input{sections/quantum-simulation}
\input{sections/quantum-ML}
\input{sections/quantum-pheno}
\input{sections/conclusions}

\begin{acknowledgments}
We would like to thank Tao Han, Henry Lamm, Hideki Okawa and Yue Zhao for their insightful feedback and valuable comments on our manuscript. Y. Fang is supported by NSFC Basic Science Centre Program for “Joint Research on High Energy Frontier Particle Physics” (Grant No. 12188102); and is supported by the Institute of High Energy Physics, Chinese Academy of Sciences under the innovative project on sciences and technologies with Grant No. E3545BU210. Y.-Y. Li is supported by the NSF of China through Grant No. 12305107, No. 12247103. J. Shu is supported by Peking University under startup Grant No. 7101302974 and the NSF of China under Grants No. 12025507, No.12150015; and is supported by the Key Research Program of Frontier Science of the Chinese Academy of Sciences under Grants No. ZDBS-LY-7003. Y. Wu and L. Xu are supported by the NSF of China International Cooperation Grant No. 12061141004, NSF of China Grant No. 12275265, and MOST National Key Technologies R\&D Program No. 2023YFA1605703. H. Xing is supported by the NSF of China with Grant No. 12035007 and Guangdong Major Project of Basic and Applied Basic Research with No. 22020B0301030008, 2022A1515010683. C. Zhou is supported by the Fundamental Research Funds for the Central Universities, Peking University.
\end{acknowledgments}

\bibliography{main}


\end{document}

%% file: sections/quantum-sensing.tex
\section{Quantum Sensing}
\label{sec:sensing}
Quantum sensors offer unprecedented accuracy in measuring physical quantities, enabling us to determine fundamental constants with greater precision and detect subtle effects predicted by BSM physics. This enhanced capability pushes the boundaries of our understanding of fundamental physics. Below, we delve into how quantum sensing can benefit three key areas: dark matter searches, tests of spacetime symmetries, and gravitational wave detection. 

\subsection*{Dark matter searches}

\textit{Wavelike dark matter --- }
The nature of Dark Matter (DM) hints at BSM physics. 
Although there is significant evidence for the existence of DM based on its gravitational effects \cite{Rubin:1980zd}, its direct detection has been elusive, and its intrinsic properties remain a mystery. 
Bosonic DM candidates in the sub-eV mass range \cite{Arias:2012az, jackson2023search} such as the QCD axions~\cite{Preskill:1982cy, Abbott:1982af, Dine:1982ah}, axion-like particles (ALPs)~\cite{Graham:2015cka, Co:2020xlh} and hidden sector photons~\cite{holdom1986two, Nelson:2011sf, Graham:2015rva}, have garnered increased attention in recent years. Their large de Broglie wavelength and high local occupation number found in DM halos suggest a wave-like nature, implying that they can be treated as coherently oscillating classical fields within the correlation time and distance.

Quantum technologies are employed to search for such wave-like DM, facilitating unprecedented precision in sensing capabilities. 
Resonant detectors such as microwave cavities or superconducting circuits are especially well-suited for searching for these forms of DM by matching the resonant frequency to their mass. 
Axions in the presence of a strong background magnetic field and dark photons, can induce effective currents through their couplings to the electromagnetic fields.
Since the experiment sensitivity increases with the quality factor of the resonant detector, a superconducting radio-frequency (SRF) cavity with an exceptionally high-quality factor $Q>10^{10}$ has generated a lot of interest in DM searches \cite{Berlin:2019ahk}. In particular, it can be shown that the scan rate of such a haloscope experiment is proportional to the loaded
quality factor even if the cavity bandwidth is much narrower than the DM halo line shape~\cite{PhysRevD.110.043022}.
For example, Ref.~\cite{SHANHE:2023kxz} has conducted a tunable SRF cavity-based experiment looking for dark photon DM at around 1.3 GHz, placing the world's leading constraint on the dark photon-photon mixing parameter for this mass range. Quantum sensors such as superconducting qubit \cite{Chen:2022quj}, trapped electrons \cite{Fan:2022uwu, fan2024ex}, Rydberg atoms \cite{Engelhardt:2023qjf} and NV centers \cite{Chigusa:2023hms}, also offer powerful tools for the direct detection of wave-like DM.

While at lower frequencies, quantum fluctuations added by the linear amplifiers employed in the state-of-the-art haloscopes establish a Standard Quantum Limit (SQL) on the detection sensitivity. Squeezed states \cite{HAYSTAC:2020kwv}  have been utilized as an attempt to go beyond the SQL, speeding up the DM search by a factor of two. The ultimate sensitivity beyond the SQL is expected to be achieved by utilizing single microwave photon detectors~\cite{Schuster:2007fwf, Johnson:2010xam}, through quantum non-demolition (QND) measurements~\cite{RevModPhys.52.341, RevModPhys.68.1}. As recently demonstrated in \cite{Dixit:2020ymh}, single microwave photon detectors were implemented by coupling the cavity to a superconducting qubit, reducing noise to 15.7 dB below the SQL limit.

\begin{figure}
    \centering
    \includegraphics[width=\linewidth]{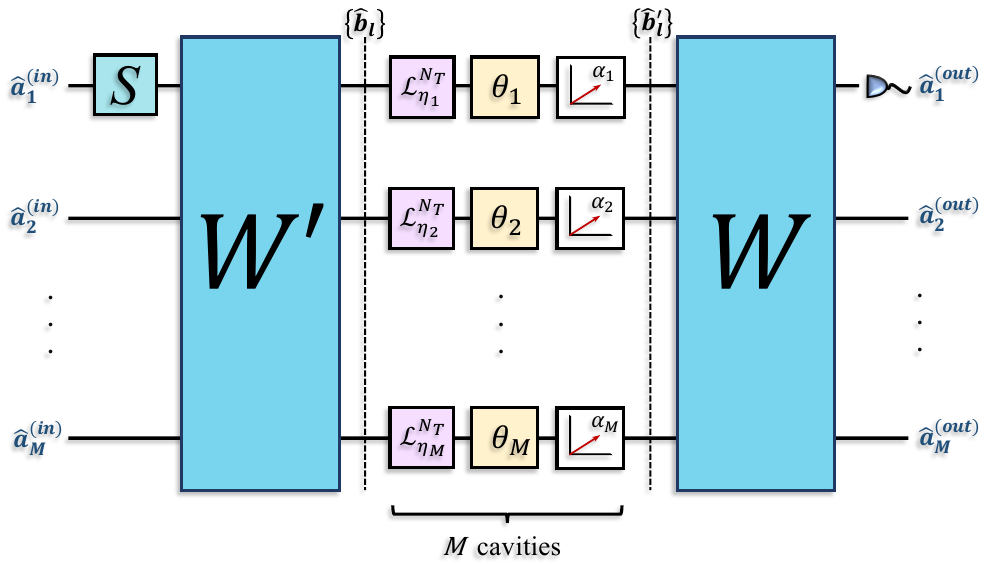}
    \caption{DQS setup: A single-mode squeezed vacuum is distributed to an array of $M$ cavities which are coupled via passive linear networks, $\bm{W}^\prime$ and $\bm{W}$. The network utilizes classical correlations between the axion-induced displacements at each cavity to coherently combine the signal fields into the primary output mode, $\hat{a}_1^{\rm (out)}$, thereby generating a larger signal amplitude. From Ref.~\cite{Brady:2022bus}.}
    \label{fig:dqs}
\end{figure}

\begin{figure}
    \centering
    \includegraphics[width=\linewidth]{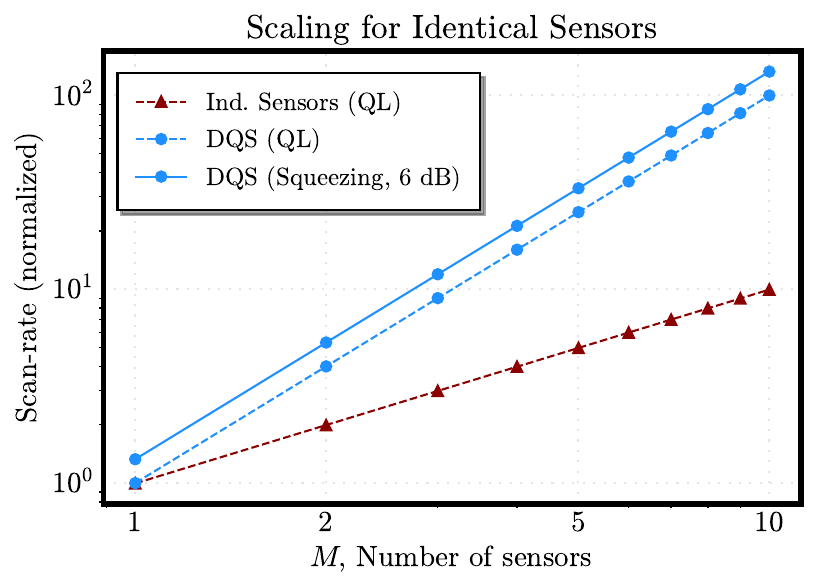}
    \caption{Scaling of the scan-rate with the number of sensors $M$ (log-log plot). Solid line corresponds to a squeezed input for a DQS setup, with gain $G\approx1.56$ (6dB of squeezing). Dashed lines correspond to QL setups with joint post-processing (QL-DCS; circles) and without joint post-processing (Ind. Sensors; triangles). Observe quadratic scaling of the scan-rate for distributed sensing scenarios versus linear scaling for independent sensors as well as a constant factor improvement for all $M$ in the DQS setup due to squeezing with fixed gain. Normalization performed is with respect to the single-cavity, quantum-limited setup. From Ref.~\cite{Brady:2022bus}.}
    \label{fig:scanrate}
\end{figure}

Quantum metrology provides various techniques to further improve the detection sensitivity and scan rate of DM detection. We may achieve Distributed Quantum Sensing (DQS)~\cite{Zhuang:2018wqs, Zhang:2020skj} and quantum-enhanced parameter estimation~\cite{giovannetti2004quantum} using a quantum network with entangled sensors. In this way, we can use $M$ entangled quantum sensors to achieve an $O(M^2)$-enhancement in scan rate with a proper set of quantum operations \cite{Chen:2021bgy, Brady:2022bus, Brady:2022qne, Chen:2023ryb, Chen:2023swh, Ito:2023zhp, Shu:2024nmc}. 
Figure \ref{fig:dqs} shows a DQS setup for cavity-based DM searches, with the dependence of its scan rate on the number of cavities depicted in Figure \ref{fig:scanrate}. 
A non-trivial initial state in the detector with a high occupation number can also increase the signal rate by stimulated emission \cite{Agrawal:2023umy}.

Numerous other techniques exist for identifying ALPs, such as the axion-induced birefringence effect for light propagation \cite{Carroll:1989vb,Harari:1992ea,Chen:2019fsq}, and the axion-induced nuclear magnetic precession \cite{Graham:2013gfa,Budker:2013hfa,Jiang:2021dby} if they are coupled to quarks or gluons.
The DM wave could also induce oscillations in the electron mass and the fine structure constant due to its coupling to the QED sector, which can then be detected by 
high precision atomic clocks \cite{Kennedy:2020bac}. 
Another effect of an oscillating QCD axion is that it could lead to oscillations in the nuclear charge radius \cite{Banerjee:2023bjc}, which would then induce oscillations in 
electronic energy levels and could be resolved with optical atomic clocks. 
Furthermore, the QCD axion could induce oscillations in the pion mass, thereby altering the proton mass and the $g$-factor, which can then be probed by atomic and molecular spectroscopy as well as interferometry experiments \cite{Kim:2022ype}. 

\textit{Particle dark matter --- } 
Traditional methods for detecting particle-like DM rely on measuring ionization signals where the DM particle scatters off an individual nucleus or electron. The typical energy threshold is about a keV, corresponding to nuclear recoils with a DM particle mass above a GeV~\cite{LZ:2022lsv}, or to MeV-scale masses for the Migdal effect~\cite{Vergados:2005dpd, Moustakidis:2005gx, Bernabei:2007jz, Ibe:2017yqa, XENON:2019zpr} or electron recoils~\cite{Tiffenberg:2017aac, SENSEI:2020dpa, XENON:2019gfn}. To detect DM particles with masses lower than the ionization threshold, new detectors, including superconductors \cite{Hochberg:2015pha, Hochberg:2015fth, Hochberg:2021pkt, Hochberg:2021ymx, Hochberg:2019cyy, Chiles:2021gxk}, superfluids \cite{Schutz:2016tid, Knapen:2016cue, Caputo:2019cyg}, polar crystals \cite{Griffin:2018bjn, Knapen:2017ekk, Cox:2019cod}, topological materials \cite{Sanchez-Martinez:2019bac}, and Dirac materials~\cite{Hochberg:2017wce, Geilhufe:2018gry, Geilhufe:2019ndy, Coskuner:2019odd}, are being developed to leverage their lower energy excitations, such as phonons, rotons, magnons, and quasiparticles (QPs) in superconductors. Superconductors show exceptional promise, as their energy threshold, set by the Cooper pair binding energy, is as low as a few meV, allowing the probing of light DM particles with sub-MeV masses. Moreover, phonons, rotons, and magnons can be collected in superconductors and converted into QPs. Thus, measuring the QPs in superconductors plays a crucial role in low-mass DM particle searches.

The electrical properties of a superconductor are determined by the density of Cooper pairs and QPs within it. To measure the Cooper pairs, one can monitor the change in electrical resistance during the superconducting transition, typically using a Transition Edge Sensor (TES), which currently offers the best energy resolution. Quasiparticle density, on the other hand, can be measured by observing variations in the inductance of the superconductor with a Kinetic Inductance Device (KID), or by detecting changes in the capacitance of a Cooper pair box coupled to the superconductor using a quantum capacitance detector.
In addition to the  linear devices mentioned above, Superconducting Nanowire Single-Photon Detectors (SNSPDs) operate in the non-linear regime and function with nearly zero dark counts at temperatures above 1 K.

Transmon qubits \cite{Riste:2013zqw} are also well-suited for QP measurements and can be used for light DM particle detection. 
Benefiting from the rapid development of quantum computing, transmon qubits have emerged as promising devices for low-threshold DM detection. Their design specifically reduces sensitivity to charge noise and flux noise through the optimization of circuit parameters, while maintaining sufficient coupling to signals of interest.  
The presence of QPs can be detected by continuously monitoring the spatial and temporal correlations of charge-parity jumps or bit flips within a multi-qubit array \cite{McEwen:2021wdg, Bratrud:2024qnk, Li:2024nhg}.

\subsection*{Tests of spacetime symmetries} 

\textit{Electric dipole moments and CP violation --- } The electric dipole moment (EDM) of a fermion $f$ is defined through $\mathcal{L}_{\rm EDM}=-\frac{i}2d_f\bar f\sigma_{\mu\nu}\gamma_5f F^{\mu\nu}$, where $F^{\mu\nu}$ is the photon field, $d_f$ is conventionally expressed in $e$~cm. In the non-relativistic limit, this interaction becomes $d_f\chi_f^\dagger\vec{\sigma}\chi_f\cdot \vec{E}$, 
and is manifestly odd under time reversal (T) symmetry. Therefore, the observation of a non-zero EDM in non-degenerate systems, such as elementary particles, atoms, or molecules would imply CP violation under the assumption of CPT invariance.
SM provides three sources of CP violations: the CP-violating phase of the Cabibbo-Kobayashi-Maskawa (CKM) matrix~\cite{Kobayashi:1973fv}, the strong CP phase $\bar\theta$, and the CP phase in the neutrino mixing matrix. 
Table~\ref{tab:edm} lists the expected contributions to EDMs from the CKM and $\bar\theta$, as well as the most stringent experimental limits on those EDMs \cite{Workman:2022ynf}. Due to the smallness of the CKM contributions, EDMs, not only severely constrain the strong CP phase ($\bar\theta\sim 10^{-10}$), but also provide a clean probe of new sources of CP violation arising from BSM physics. 
In recent years, atom- and molecule-based searches for CP violations have advanced rapidly, and have benefited from the advancement of QIS. Looking ahead, next-generation experiments employing laser cooling and trapping at ultra-cold temperatures \cite{PhysRevLett.121.013202, PhysRevLett.123.033202, Anderegg_2019, PhysRevX.10.021049, FITCH2021157}, matrix isolation \cite{atoms6010003,PhysRevA.98.032513,Singh2019,PhysRevLett.125.043601}, advanced quantum control \cite{Cloet:2019wre, Hosten2016, Aoki_2021} aim to improve precision by a factor of $10^4$ for the electron EDM \cite{Alarcon:2022ero}. 
This could potentially uncover new sources of CP violation and provide insights into BSM physics.

\begin{table}[h!]
\centering
\begin{tabular}{|l |c | c |l|l|} 
 \hline
EDM & ${|d_f|}_{\rm CKM}$ exp. & ${|d_f|}_{\bar\theta}$ exp. & Limit ($e$~cm) & Source\\ [0.5ex] 
 \hline
 Electron & $ 10^{-44}$ & &$4.1\times 10^{-30}$ &  HfF$^+$\cite{Roussy:2022cmp}\\ 
 &  & &$1.1\times 10^{-29}$ & ThO\cite{Andreev2018} \\ 
 Neutron & $(1-6)\times 10^{-32}$ & $
 10^{-16}\bar\theta $ &$1.8\times 10^{-26}$  &\cite{Abel:2020pzs} \\
 $^{199}$Hg  & $\lesssim 4\times 10^{-34}$ & $
 10^{-20}\bar\theta $ & $7.4\times 10^{-30}$  &\cite{PhysRevLett.116.161601}\\
 \hline
\end{tabular}
\caption{The most stringent experimental limits on EDMs are given, along with the theoretical expectations of the CKM matrix and/or the strong CP phase's contributions to the EDMs of the electron \cite{Pospelov:2005pr}, neutron \cite{PhysRevC.91.025502} and  $^{199}$Hg \cite{Chupp:2017rkp}. }
\label{tab:edm}
\end{table}

\textit{Lorentz and CPT symmetry --- }
CPT is an exact symmetry of any local and Lorentz-invariant quantum field theory (QFT) with a positive-definite hermitian Hamiltonian that preserves causality~\cite{Workman:2022ynf}.
A violation of CPT implies either a violation of Lorentz invariance, locality, or unitarity~\cite{PhysRevLett.89.231602}. 
Under the CPT symmetry, a particle and its antiparticle must have equal masses and widths. Based on this, experimental tests of Lorentz and CPT symmetry violation often employ precision comparisons of particles and antiparticles, such as hydrogen vs. anti-hydrogen spectroscopy \cite{Ahmadi2018}, studies of neutral kaon systems \cite{CPLEAR:1999ysk}, and Penning trap measurements of proton and antiproton \cite{Borchert2022}. Concurrently, atomic clocks, atom magnetometry, and other precision spectroscopy experiments have yielded some of the most stringent bounds on  Lorentz violation \cite{PhysRevLett.111.050401,PhysRevLett.122.123605,Sanner2019,Pruttivarasin2015}.  
Lately, the rapid development of QIS has the potential to further tighten constraints or possibly detect tiny violations in future experiments, through improvements in clock precision \cite{RevModPhys.90.045005,PhysRevLett.120.103202}, quantum-logic-based spin readout \cite{Nitzschke:2019roc}, and many other techniques \cite{Adelberger:2022sve}. The continued scrutiny is important for HEP, as any observed Lorentz/CPT violation would have profound implications for our understanding of nature at the most fundamental level. \\

\textit{Equivalence Principle --- } One of the defining postulates of Einstein’s theory of General Relativity (GR) concerns the interaction between matter and gravity. It states that the metric tensor couples universally and minimally to all the SM fields, which leads to many observable consequences that go under the ``Equivalence Principle (EP)", such as
\begin{itemize}
\item Universality of free fall or Weak Equivalence Principle (WEP). The Strong Equivalence Principle extends WEP to self-gravitating bodies.
 \item Einstein Equivalence Principle (EEP), which implies constancy of constants (e.g., fine structure constant $\alpha$) and local Lorentz invariance.
\item Universal gravitational redshift of clock rates.
\end{itemize}
Table~\ref{tab:ep} lists some of the latest results from various EP tests \cite{Workman:2022ynf}. Universality of free fall has been tested to a precision of one part in $10^{15} (10^{13})$ in space-based (ground-based) experiments, with 100 times improvement in projected sensitivity using quantum superposition of cold atoms in space~\cite{STE-QUEST:2022eww}. Comparisons of Yb$^+$, Sr clock transition frequencies not only constrained the time variation of the fine structure constant and the proton-electron mass ratio to approximately one part in $10^{18}$ per year, but also provided the most stringent tests of local Lorentz invariance \cite{Sanner_2019}. Using atomic clock network \cite{Barontini:2021mvu} or space-based atomic clocks \cite{Schkolnik:2022utn} can further improve the precision in the future. 
Beyond atomic clocks, nuclear clocks using $^{229}$Th have attracted significant attention. The smallness of the nucleus and the existence of a low-lying nuclear transition which allows its manipulation by a laser of $\sim$150 nm~\cite{Kraemer:2022gpi} make $^{229}$Th unique among nuclear isotopes and promising for ultra-precise timekeeping.

\begin{table}[h!]
\centering
\resizebox{0.99\linewidth}{!}{
\begin{tabular}{|l | c |c|} 
 \hline
 Test & Method & Constraint \\ [0.5ex] 
 \hline
WEP &  Satellite based free fall  & grav. acc. $\Delta a/a\sim 10^{-15}$\cite{PhysRevLett.129.121102} \\
& Torsion-balance test  &  $\Delta a/a\sim 10^{-13} $ \cite{Wagner_2012}\\
\hline
EEP & Atomic clocks \cite{Lange_2021, Filzinger_2023} & $\dot\alpha/\alpha\sim(1.8 \pm 2.5) \times 10^{-19}{\rm yr}^{-1}$ \\ 
 &  $\alpha,~ \mu(= m_p/m_e)$  & $\dot\mu/\mu\sim(-8 \pm 36) \times 10^{-18}{\rm yr}^{-1}$ \\ \hline
\end{tabular}}
\caption{Examples of some of the WEP and EEP tests, their methods and the corresponding constraints.}
\label{tab:ep}
\end{table}

\subsection*{Gravitational wave detection}

\textit{Gravitational wave astronomy --- } opens up a unique window into phenomena that remain invisible to traditional electromagnetic observations. This groundbreaking field enables us to study black holes and neutron stars in unprecedented detail, providing crucial insights into their formation, evolution, and distribution throughout cosmic history. Moreover, gravitational waves serve as a powerful tool for testing GR as well as offering an independent method for measuring the expansion rate of the universe~\cite{LIGOScientific:2017adf}. Additionally, this new observational technique can also be applied to DM searches.

Since the first detection in 2015 \cite{LIGOScientific:2016aoc}, the ground-based observatories LIGO, Virgo, and KAGRA have observed over 90 gravitational wave events~\cite{KAGRA:2021vkt}, predominantly from binary black hole mergers, but also including binary neutron star and neutron star-black hole systems~\cite{LIGOScientific:2017vwq,LIGOScientific:2020aai,LIGOScientific:2021qlt}. 
Quantum sensing has significantly improved the sensitivity of current detectors. 
Advanced LIGO and other detectors now routinely use squeezed quantum states of light~\cite{Miller:2014kma} to reduce quantum noise, particularly shot noise at high frequencies. 
Additionally, the large mirrors in LIGO, weighing 40 kg, have been shown to be entangled by the squeezed states of light in the interferometer, demonstrating the reach of quantum effects at truly macroscopic scales~\cite{LIGOScientific:2020luc}. 
The field is also expanding beyond ground-based interferometers. Space-based detectors such as LISA~\cite{amaroseoane2017laserinterferometerspaceantenna}, Taiji \cite{taiji} and TianQin \cite{Luo_2016} are under development, aiming to access lower frequency gravitational waves. Meanwhile, atom interferometry~\cite{Abe_2021, Canuel_2018,Zhan_2019,AEDGE:2019nxb,Canuel:2019abg,Badurina:2019hst} is being explored to bridge the mid-frequency gap between ground and space-based detectors.

Looking to the future, further improvements in the optical quantum measurement techniques are being developed that allow for simultaneous suppression of radiation pressure and shot noise~\cite{ballmer2022snowmass2021cosmicfrontierwhite}. 
Further advancements in quantum-enhanced metrology could yield measurement precisions beyond the SQL. For instance, spin squeezing~\cite{Greve:2021wil}, particularly relevant for atom interferometers, offers a promising way to surpass this limit. These quantum technologies present tantalizing possibilities for the future of gravitational wave astronomy. 
However, implementing these advanced quantum techniques in large-scale detectors presents significant technical challenges. 

Above 10 kHz, which is beyond the capabilities of current gravitational wave detectors, new detection schemes are required. Quantum sensing can also play a major role here. Ref.~\cite{PhysRevLett.110.071105,PhysRevLett.128.111101} proposed using optically levitated sensors as ultra-sensitive force detectors to search for gravitational waves up to hundreds of kHz. Ref.~    \cite{Berlin:2021txa} proposed to use superconducting cavities to detect the inverse Gertsenshtein effect~\cite{Boccaletti:1970pxw} in the 1-10 GHz range.
While these quantum sensing approaches show great promise for high-frequency gravitational wave detection, significant technological development is still needed to reach the sensitivities required to detect predicted sources in this regime. The combination of quantum sensing techniques with innovative detector designs specifically tailored for high frequencies represents a promising direction for expanding the frontier of gravitational wave astronomy into the MHz-GHz band. \\ 

%% file: sections/quantum-simulation.tex
  \section{Quantum Simulation}
\label{sec:simulation} 
The dynamical properties of matter in HEP encompass diverse topics, including non-equilibrium dynamics in the early universe, non-perturbative real-time dynamics that involve parton distribution functions (PDFs) and fragmentation functions (FFs), the phases of matter in heavy ion collisions and in neutron stars, and coherent processes in neutrino physics.
All of these stem from the nature of quantum fields upon which our theoretical framework is built. First-principle calculations of these dynamics with classical computers require resources that are exponential functions of the system size, due to the notorious sign problem. Recent experimental advancements in controlling quantum many-body systems, such as those described in \cite{3DHubbard}, have ushered in a new era of computational paradigms: quantum computing. Though quantum devices are known to be susceptible to errors due to the delicate nature of quantum states, both hardware technologies and theoretical studies in quantum error correction are advancing (see, e.g., \cite{proctor2024, IBM-QEC}) to push the error rates of quantum devices below the thresholds \cite{shor1997faulttolerant}, where reliable large-scale (fault-tolerant) quantum simulations can be achieved. By efficiently exploring vast Hilbert spaces and simulating local Hamiltonians \cite{Lloyd:1996aai}, quantum computing offers advantages for performing first-principles calculations of these dynamics in polynomial time. 
Advances in this direction require lines of research from theory to experiment \cite{Catterall:2022wjq, Beck:2023xhh, PRXQuantum.4.027001, np-review, Di_Meglio_2024}, including developments of general quantum algorithms to simulate dynamics of QFT non-perturbatively, as well as benchmark studies in the noisy intermediate-scale quantum (NISQ) era that could provide guidelines for our understandings of the nature and also hardware developments. For explorations of quantum algorithms in perturbative calculations, see e.g. \cite{Ramirez-Uribe:2021ubp,clemente2023, delejarza2024}.

\subsection*{Quantum algorithms for QFT}
After Feynman posed the question of simulating QFT on a quantum computer \cite{Feynman:1981tf}, quantum algorithms for non-perturbative calculations have been under development, starting with simulations of the dynamics of quantum systems \cite{Lloyd:1996aai} with a fixed number of particles, extending to lattice gauge theories (LGTs) \cite{PhysRevA.73.022328}, and later, by Jordan, Lee, and Preskill (JLP) \cite{Jordan:2012xnu}, addressing the convergence to the continuum limit. Algorithms developed so far often involve two essential truncations: discretizing the space and digitizing the bosonic field, which allow for mapping the field’s infinite degrees of freedom (DoF) onto a finite number of qubits. They further require preparations of initial states from the vacuum, the real-time evolution of the system, and the measurement of observables. Theoretical studies are emerging that investigate the systematic errors from these truncations and approximations used in the real-time evolution, as well as develop efficient methods for achieving the continuum limit. 

\textit{Discretization -- }The first essential truncation is to discretize the space on lattices. Other discretization methods on lattices that differ from spatial lattices can be found in, e.g. \cite{e23050597, PhysRevA.103.062601, PhysRevA.105.032418, Buser_2021}. The gauge-invariant Hamiltonian to describe the gauge theories on a lattice was developed by John Kogut and Leonard Susskind \cite{PhysRevD.11.395, PhysRevD.13.1043} in the temporal gauge $A_0=0$, in which fermions are placed on lattice sites and gauge bosons on the links between sites. This is the well-known KS Hamiltonian, with errors to the Hamiltonian in the continuous space of $\mathcal{O}(a)$, where $a$ is the lattice spacing. The KS Hamiltonian is most commonly considered in quantum simulations at the present time. Including more gauge-invariant terms in the Hamiltonian can improve the convergence rate to the continuous-space theory by pushing the discretization errors to higher orders of $a$ \cite{PhysRevD.64.094503, PhysRevD.59.034503, PhysRevD.59.014511, PhysRevD.59.074502, PhysRevD.75.054502, LEPAGE1998267}. Quantum simulations of these improved Hamiltonians have recently been investigated \cite{Carena:2022kpg, PhysRevD.108.094513, Gustafson:2023aai} with the primary goal of reducing the resource consumption required in approaching the continuum limit. Nonetheless, more effort is needed to quantify improvements in the convergence rate using non-perturbative methods.

\textit{Digitization -- }The second essential truncation involves digitizing the continuous field DoF into finite ones to allow their mapping onto qubits. Since the temporal gauge in the Hamiltonian formulation doesn't completely fix the gauge, the full Hilbert space spanned by all links on the lattice contains redundant degrees of freedom connected via gauge transformations. These redundancies can be eliminated by projecting onto the gauge-invariant subspace through the imposition of Gauss's law constraints. These two different Hilbert spaces naturally lead to two classes of digitization, with various methods being proposed for each. One class digitizes the full Hilbert space that includes discrete subgroups in the magnetic basis \cite{Ji_2020, Carena:2021ltu, PhysRevLett.129.051601,Gustafson:2023kvd, Ji_2023}, a finite number of irreducible representations or Schwinger bosons in the electric basis \cite{Zohar:2012xf,Zohar:2012ay,Zohar:2014qma,Zohar:2015hwa,Singh:2019uwd,Singh:2019jog,Buser:2020uzs}, fuzzy gauge theory \cite{Alexandru:2022son,Alexandru:2023qzd}, quantum link models \cite{Wiese:2014rla,Luo:2019vmi,Brower:2020huh,Mathis:2020fuo,Halimeh:2023lid,Halimeh_2024}. To perform real-time evolutions, quantum algorithms need to be implemented to prepare gauge-invariant initial states out of the redundant Hilbert space. Such algorithms have recently been realized for digitization methods in the magnetic basis \cite{Carena:2024dzu}, while await to be explored for other digitizations. The second class digitizes only the gauge-invariant states, which includes digitizing the plaquettes \cite{Zohar:2013zla,Kaplan18_GaussLaw,Bender20_compactQED,Yamamoto:2020eqi,Bauer:2021gek,Grabowska:2022uos,Kane:2022ejm,PRXQuantum.2.030334, Ciavarella:2024fzw}, states outside the maximal tree in the maximal gauge \cite{Bauer:2023jvw}, gauge-invariant bosonic and fermionic operators in the loop-string-hadron (LSH) formulations \cite{Anishetty_2009, PhysRevD.90.114503, PhysRevD.101.114502}, Fock states in the light-front Hamiltonian \cite{Barata_2022, Barata_2023,  Kreshchuk:2020kcz,Kreshchuk:2020dla,lfns:2024, qian2024}, local multiplet states \cite{Klco:2019evd,Ciavarella:2021nmj},
and others\cite{Haase2021resourceefficient, Zache:2023dko,Hayata:2023bgh, Li:2024nod, fontana2024}.

Digitizing gauge-invariant states offers advantages in reducing the qubit requirement and avoiding the step of preparing gauge-invariant initial states. However, this process can complicate the Hamiltonian by introducing non-local interactions upon implementing Gauss's law constraints. This non-locality can naively lead to exponential scaling of the gate count with the volume. To address this issue, operator redefinitions have been employed for compact U(1) gauge theories \cite{Kane:2022ejm, Kane:2022rlr}, which effectively reduce the non-locality and break the exponential scaling. Nevertheless, the persistence of exponential scaling remains a challenge for non-Abelian gauge theories. On the other hand, when digitizing the full Hilbert space, the Hamiltonian remains local. The inherent redundancies within the Hilbert space have been leveraged to make strides in quantum error correction and mitigation \cite{Stannigel:2013zka, Halimeh_2020, PhysRevLett.125.240405, Halimeh_2021,Bonati:2021vvs,Bonati:2021hzo,Gustafson:2023swx,Tran:2020azk,Lamm:2020jwv,Stryker:2018efp,Rajput2023npj,Mathew:2022nep,VanDamme:2021njp,Halimeh:2022mct}. With the correctable errors identified, error thresholds can be derived \cite{Carena:2024dzu} below which keeping gauge redundancies for quantum error corrections can achieve higher fidelity than removing these redundancies, providing important guidance for fault-tolerant quantum
simulations of gauge theories.

Though with these rapid advancements, we have yet to reach a stage where it becomes feasible to comprehensively compare various digitization methods in terms of resource demands, convergence rate to the continuous theory, quantum fidelities in the NISQ era, and other relevant aspects. Recently, there have been notable estimates of resource requirements for discrete group methods \cite{Gustafson:2023kvd, Gustafson:2024kym, Lamm:2024jnl}, the LSH formalism, and the Schwinger boson approach \cite{Davoudi_2023}, alongside others \cite{PhysRevD.102.094501, PhysRevA.105.032418}, when considering qubit-based quantum computing platforms. In \cite{Lamm:2024jnl}, a comparative analysis of resource requirements in terms of T-gate counts was conducted across different digitization methods for $SU(2)$ gauge theory in 1+1D for a single-step time evolution of the KS Hamiltonian. Given the costly encoding required for error correction associated with T gates, this metric for resource demand is of particular interest. It turns out that the T gate counts, when mapping $SU(2)$ gauge bosons to 7 qubits, exhibit similar magnitudes \cite{Lamm:2024jnl, Davoudi_2023}. Consequently, it is imperative to rigorously quantify both systematic errors inherent to these digitization strategies. Moreover, an evaluation of the circuit depth necessitates a side-by-side comparison between different approaches for approximating the time evolution operators in LGTs \cite{Shaw:2020udc, Rhodes:2024zbr}.

The field DoF can also be mapped onto multi-level qudit systems using various digitization methods \cite{Gustafson:2021qbt, 9651438, PhysRevLett.129.160501, kurkcuoglu2022quantumsimulationphi4theories, Zache2023fermionquditquantum, PhysRevResearch.6.013202, Calajo:2024qrc, Illa_2024, murairi2024, doga2024qudit}, which have the potential to reduce the required circuit depth and thus gate errors. Encoding gauge field DoF into bosonic DoF, such as phonons \cite{Davoudi:2021ney}, has also been studied to simulate the dynamics of a finite number of bosonic excitations. Quantum computation via quantum optics \cite{Marshall:2015mna, Abel:2024kuv}, by offering an infinite-dimensional photon-number DoF, can preserve the continuous nature of boson fields and promise accurate simulations of QFT. 

\begin{figure*}[htbp]
    \centering
\begin{minipage}[h]{0.67\linewidth}
      \vspace{0pt}
\includegraphics[width=\linewidth]{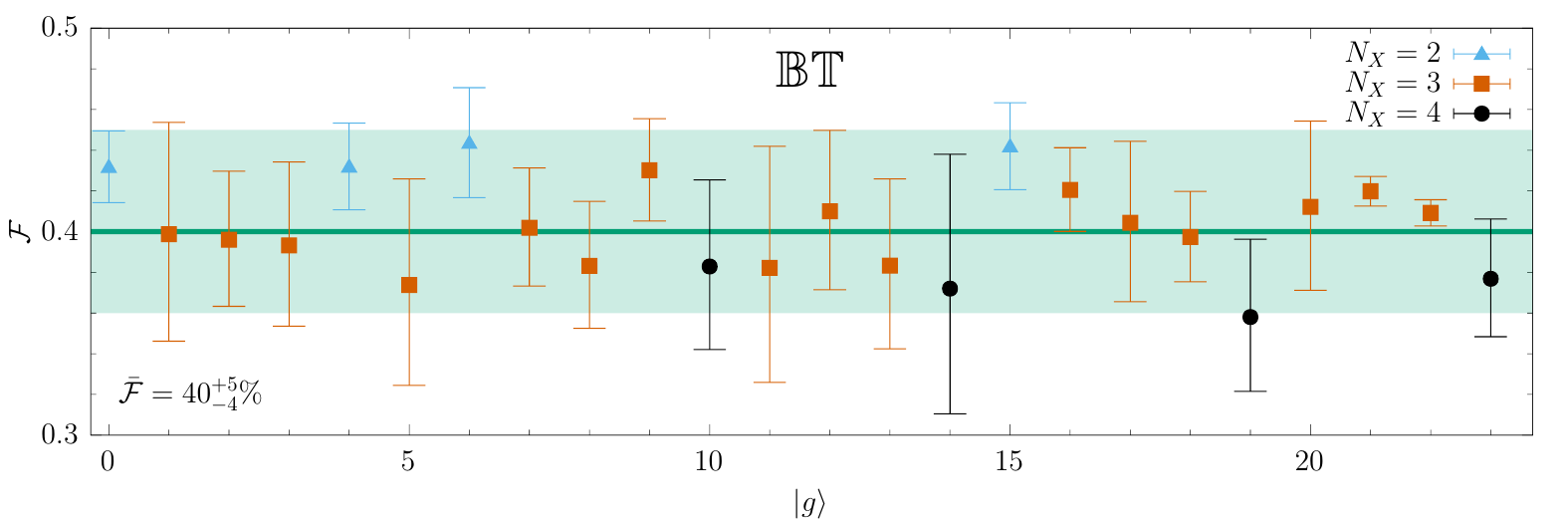} 
     \end{minipage}
     ~\begin{minipage}[h]{0.15\linewidth}
      \vspace{-10pt}      \includegraphics[width=\linewidth]{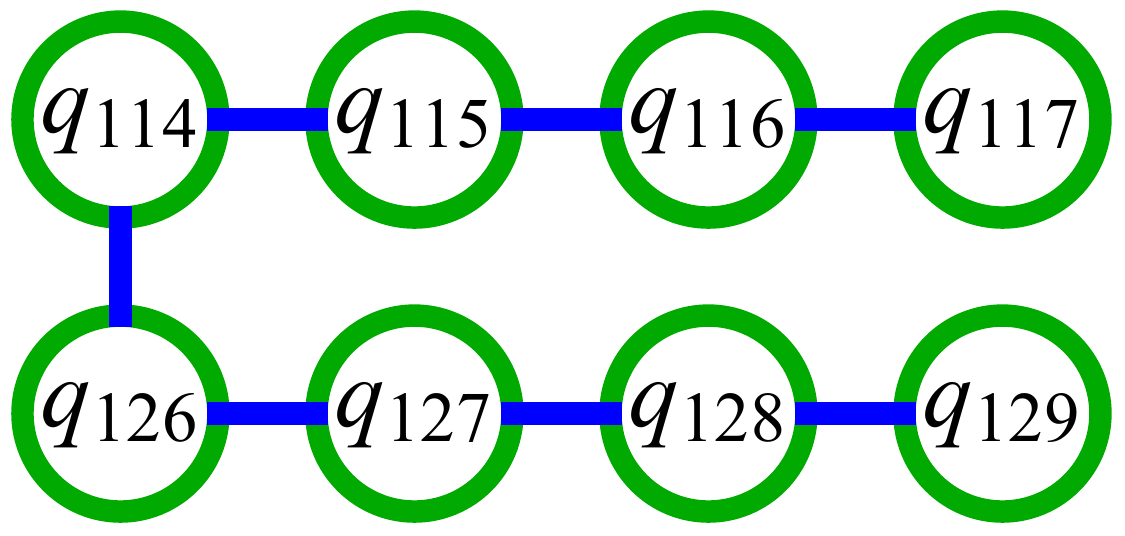} 
     \end{minipage}
    \caption{Fidelity of $\mathfrak{U}^{\rm trans}_{-1}$ gate of \texttt{Baiwang} for each group element $|g\rangle = |abcd\rangle$ of $\btt$ \cite{Lamm:2024jnl}, labeled by the lexicographic order in the range $\{0,|G|-1\}$ with $|G|$ being the dimension of the group . The averaged fidelity $\bar{\mathcal{F}}$ over all group elements are also shown. On the top right, the qubit graph on \texttt{Baiwang} used to represent $\ket{g}$ for $\btt$ are shown.}
    \label{fig:fidelity}
\end{figure*}
We provide an illustrative example of simulating the dynamics of $SU(2)$ gauge theories using the discrete subgroup digitization method. $SU(2)$ gauge theory is represented by one of its discrete subgroup $\btt$. The group element of $\btt$ can be encoded into qubits using either the ordered product methods \cite{Gustafson:2022xdt, Gustafson:2023kvd} or the block encoding method \cite{Lamm:2024jnl} with the latter being applicable to the largest crystal-like subgroup of $SU(2)$. Adopting the block encoding, each group element is mapped to $n_q=8$ qubits where redundant degrees of freedom are removed by imposing additional constraints with quantum circuits. The 8 qubits shown in the right panel of \fig{fidelity} are chosen using the \texttt{Baiwang} quantum real machine on Quafu \cite{bib40} for this purpose. Starting from the state $|\psi_0\rangle = |0\rangle^{\otimes n_q}$, we can prepare the initial state $|g\rangle$, representing the desired group element $g$, with circuit $\mathfrak{U}_g$: $|g\rangle = \mathfrak{U}_g |\psi_0\rangle$. One can find that $\mathfrak{U}_g$ is simply a tensor product of X gates used to initialize some qubits to the $|1\rangle$ state. We focus on simulating one specific primary gate -- the inverse operation $\mathfrak{U}_{-1}$ of the group element, leveraging the fact that the full time evolution of the KS Hamiltonian can be constructed by combining various primary operations of this type on group elements. With the connectivity of the 8 qubits on \texttt{Baiwang} quantum chip, $\mathfrak{U}_{-1}$ requiring four swap gates, is transpiled to $\mathfrak{U}^{\rm trans}_{-1}$ with 18 CNOTs. The fidelity for the inverse operation is defined as:
\begin{equation}
    \mathcal{F} = |\langle g^{-1}|\mathfrak{U}^{\rm trans}_{-1}|g\rangle|^2 =  |\langle g^{-1}|\mathfrak{U}^{\rm trans}_{-1}\mathfrak{U}_g |\psi_0\rangle|^2,
\end{equation}
which is the probability of measuring the correct $|g^{-1}\rangle$. We observe from \fig{fidelity} that the \texttt{Baiwang} quantum chip is able to simulate the one-link dynamics of the $\btt$ gauge theory with a fidelity of around 40\% across the different group elements prepared. 

The fermion field is locally of finite dimensions, and encoding the fermionic DoF on a quantum computer requires realizing the intrinsic Fermi statistics. Jordan-Wigner encoding \cite{JW-transformation} maps fermionic operators to strings of Pauli operators which leads to nearest-neighbor interactions for one-dimensional fermionic Hamiltonians. For higher dimensions, non-local qubit operators are introduced after JW transformation. Bravyi-Kitaev encoding \cite{BRAVYI2002210} allows local encodings of fermionic theories by keeping fermionic statistics at the hardware level. This fermionic quantum processor is being investigated in terms of developing protocols for fermionic gates, circuit decompositions and including coupling to gauge fields on Rydberg-based architecture \cite{fermionicprocessor, Zache:2023cfj}. Other encoding methods for fermionic DoF \cite{Whitfield_2016, PhysRevA.92.042303, doi:10.1073/pnas.0808245105} are worthy of further study in adaption to the developments of hardware. Encoding methods that take into account error correction for fault-tolerant quantum computing are also emerging \cite{landahl2023}.

\textit{Continuous limits -- } To predict quantities used as inputs for experiments, the results from quantum simulations need to be extrapolated to the limit of continuous gauge-boson fields and also to the continuous spacetime limit of infinite volume. In approaching the continuous gauge-boson field limit, a sufficient number of DoF need to be retained in the digitizations, while improved Hamiltonians may be needed for discrete subgroup digitization methods, given that only a finite number of crystal-like subgroups exist for non-Abelian gauge theories \cite{PhysRevD.100.114501}. Efforts in extrapolating to the continuous spacetime limit and understanding the systematic uncertainties from finite volume in real-time dynamics remain underdeveloped \cite{Carena:2021ltu, PhysRevD.106.114504, PhysRevD.103.014506}, where progress can be made.

{\textit{State Preparation -- }}
In QFT, the dimension of the Hilbert space grows exponentially with the system size, posing significant challenges for initial states preparation in simulating non-perturbative aspects of real-time and non-equilibrium dynamics in HEP. In the JLP protocol, the interacting wave packets are created by evolving the spatially well-separated free wave packets with a time-dependent Hamiltonian, where the interactions are turned on adiabatically. However, the adiabatic evolution strains the resources of present-day quantum hardware, limiting the application of this methodology to bound state scatterings. This challenge has motivated alternative proposals, such as directly creating wave packets in the interacting theory \cite{Davoudi:2024wyv}, employing Haag-Ruelle formalism \cite{Turco:2023rmx} and using sequency truncations \cite{Li:2024lrl} within adiabatic framework, using optical theorem \cite{Ciavarella:2020vqm}, Lehmann-Symanzik-Zimmermann reduction formula \cite{Li:2023kex}, stochastic sampling\cite{PhysRevLett.121.170501, stochastic, Harmalkar:2020mpd} and algorithms based on the
quantum eigenvalue transformation for unitary matrices \cite{Kane:2023jdo} beyond the adiabatic framework.

A more extensively used method recently is the Variational Quantum Algorithm (VQA), initially proposed in Ref. \cite{Peruzzo:2013bzg}. This approach has been successfully implemented across various disciplines of quantum physics, as reviewed in \cite{Bharti:2021zez,Cerezo:2020jpv,Tilly:2021jem}. The VQA approach takes advantage of the variational principle to search for the ground state energy of a given Hamiltonian in polynomial time scale. It involves parameterizing and optimizing a unitary quantum circuit using classical training methods, thereby reducing the quantum resources required in NISQ devices with limited numbers of qubits and quantum circuit depth. With these resource-efficient schemes for ground-state preparation, tremendous efforts have been devoted to hadronic states preparation. This includes quark-antiquark bound state preparation in the Nambu-Jona-Lasinio (NJL) model for hadron partonic structure \cite{Li:2021kcs}, hadron mass spectrum calculations including both meson and baryon states in 1+1D $SU(2)$ gauge theory \cite{Atas:2021ext}, heavy quarkonium spectroscopy studies based on a Cornell-potential model for the heavy quark and antiquark system \cite{Gallimore:2022hai,deArenaza:2024dhe}. 
Ansatz of the variational circuits are also being actively developed based on evolutions in the imaginary time and symmetries which can improve convergence to the ground state and efficiency in quantum resources required \cite{iHVA, wang2023symmetry, PhysRevLett.132.150603, Wang:2024jis}. Structure-preserved ansatz has also been developed to efficiently prepare scale-invariant ground state \cite{Li:2023vwx}.

Alternatively, Quantum Optimal Control (QOC) techniques that operate directly at the level of hardware pulses have been proposed for gate-free state preparation \cite{Meitei:2021ryz}. This approach is expected to further alleviate the constraints on coherence time in quantum hardware. QOC has been applied to show speed-up preparations for vacuum ground state in LGTs \cite{doga2024qudit} and for thermal states in the 1+1D Schwinger model \cite{Araz:2024bgg}, although practical implementations on quantum hardware remain challenging.

State preparation algorithms are being developed to expand their capability to prepare non-trivial initial states of large systems. To demonstrate the vitality of this topic, we detail a new algorithm, dubbed the SC-ADAPT-VQE (Scalable Circuits Adaptive Derivative-Assembled Pseudo-Trotter ansatz Variational Quantum Eigensolver), which was recently proposed in \cite{Farrell:2023fgd} for preparing states of arbitrarily large systems and extended in \cite{gustafson2024surrogate} to advance the simulation. By taking into account the translational invariance and the exponential decay of correlations for gapped systems, which are of broad interest, this algorithm utilizes extrapolations from circuits designed for small systems to identify scalable circuit configurations. As comprehensively studied in \cite{Farrell:2023fgd} for the $1+1$D Schwinger model, the quantum circuits for ground state preparation for systems with spatial sites $L \leq 14$ can be determined using classical computers based on an adaptive variational algorithm \cite{ada-var}. Quantum circuits for larger systems to the infinite-volume case $L = \infty$ are subsequently obtained from extrapolations of the variational parameters, which converge exponentially with $L$. One way to see the validity of this new algorithm is by measuring the chiral condensate of the state prepared and comparing it with that obtained using classical matrix product state calculations. Good agreement is found, as shown in Fig. 7 of \cite{Farrell:2023fgd}, for implementing this algorithm on the \texttt{ibm\_brisbane} quantum computer for $L=30$.

\subsection*{Benchmark studies}
In the NISQ era, benchmarks can be explored on quantum computers for algorithm development to perform measurements, as well as for understanding physics related to real-time scattering, parton distribution and fragmentation at colliders, entanglement entropy in different phases of matter at finite density, etc. Hadron dynamics are simulated using 112 qubits with the initial hadron wave-packets prepared using the SC-ADAPT-VQE algorithm \cite{Farrell:2024fit}. Phase shifts of scattering processes estimated from asymptotic state long after the collision processes, are shown to be extractable using information at an early stage of collisions \cite{PhysRevD.104.054507}, avoiding decoherence and deep circuits in the NISQ era. Quantum algorithms utilizing ancillary qubits to reconstruct $n$-time correlation functions have been proposed \cite{n-time-correlation} and have been applied to linear response theory, PDF calculation in the NJL model \cite{Li:2021kcs} on quantum platforms. Parton shower including quantum interference effects can be formulated on quantum computer based on toy models \cite{Nachman_2021, Bauer:2023ujy} and from first principles \cite{Bauer:2021gup}. Thermalization dynamics of $U(1)$ gauge theory in 1+1D has been studied on quantum computers using the quantum link models \cite{Zhou_2022}.  Accessing full final-state wavefunction is in general prohibited, new techniques are being developed to enable practical implementations on near-term quantum computers. Classical shadows \cite{shadow-measurement, shadow-review}, which construct an approximate classical description of a quantum state from randomized measurements, allow sampling of the density matrix to reduce computational resources substantially for a wide range of observables, e.g. two-point correlation functions of the ground states\cite{shadow-measurement}.

\begin{figure}[htbp]
    \centering
\begin{minipage}[h]{\linewidth}
      \vspace{0pt}
\includegraphics[width=\linewidth]{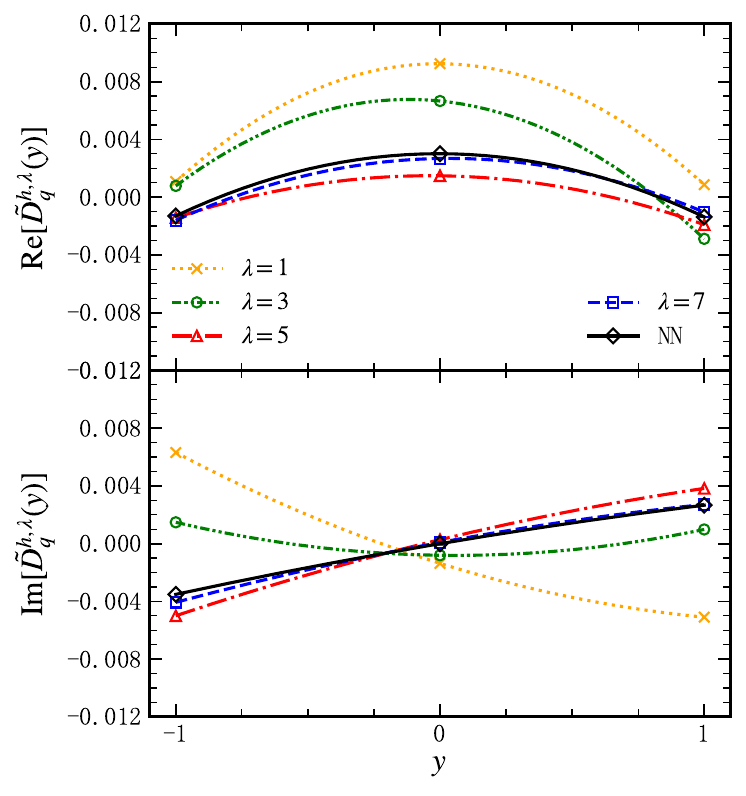} 
     \end{minipage}
    \caption{Figure taken from \cite{Li:2024nod}: Real (top panel) and imaginary (bottom panel) part of FF in position space with quantum gate fidelity 0.999, NN is the result with the zero-noise quantum circuit, $\lambda = 1$ is the noisy result after post-selection and $\lambda = k ~(k > 1)$ is the $k$-th Richardson extrapolation result after post-selection.}
    \label{fig:err-FFs}
\end{figure}

Quantum simulation of the SM requires large-scale fault-tolerant quantum computers. However, in the NISQ era, the noise in quantum operations caused by the limited qubit coherence time and gate fidelities hinders such simulations. Even for proof-of-principle experiments, error mitigation is essential to obtain reliable estimates of observables using existing hardware. The NISQ quantum circuits are primarily affected by readout errors and accumulated gate errors. Various error mitigation strategies have been proposed, and combining different error mitigation algorithms can significantly improve the signal-to-noise ratio. Fig. \ref{fig:err-FFs} serves as an example of how error mitigation can enhance the quantum simulation of FFs in NJL model \cite{Li:2024nod}, where a depolarization model is implemented to represent noise on quantum hardware.
It turns out that the impact of quantum noises cannot be ignored, as it can lead to unreliable results due to the mixing of non-physical states, and make the commonly used zero-noise extrapolation method \cite{Temme:2016vkz,9259940} fail. However, incorporating post-selection along with Richardson extrapolation demonstrates convergence to zero-noise results with a high order of extrapolation parameter $\lambda$, highlighting the effectiveness of error mitigation techniques in achieving reliable simulations of FFs.

%% file: sections/quantum-ML.tex
\section{Quantum Machine Learning}
\label{sec:qml}
HEP experiments generate data of immense complexity and volume, posing significant challenges in data analysis for the objectives of precision measurements and new physics hunt. Machine learning, as a transformative tool, has been extensively studied in HEP data analysis to enhance the overall performances regarding training efficiency and detection accuracy, see a live reviewed in \cite{Feickert:2021ajf}. As facilities like the upcoming HL-LHC and future colliders will produce more than ten times larger datasets, these classical ML methods might become increasingly strained. Conversely, quantum machine learning (QML), leveraging its parallel processing capabilities and quantum properties such as entanglement, may offer the potential to reduce computational resources as well as further improving the overall performances in HEP data analysis. Extensive work in this direction has emerged recently, exploring the potentials of QML in HEP and looking for quantum advantages in the context of speedup in training and improved performance, as reported in \cite{Delgado:2022tpc, Wu:2022icw, Guan:2020bdl}. In the following, we provide a concise summary of these studies, which encompass a wide range of data analysis tasks, including object reconstruction, quantum generative models, classification, and anomaly detection, highlighting the perspectives they offer related to quantum advantages, and discuss future directions at the end.

\textit{Object reconstruction ---}\label{sec:object-reconstruction} 
In the forthcoming HL-LHC, each collision involves a large number of particle interactions, making the reconstruction of physical objects (including tracks, jets, etc.) extremely challenging. The grouping problem is a core issue in the reconstruction of HEP experimental studies. Compared to commonly used algorithms today, quantum algorithms may offer better grouping capabilities and bring advantages in computing speed in the future.

\begin{figure}[htbp]
    \centering
    \includegraphics[width=.45\textwidth]{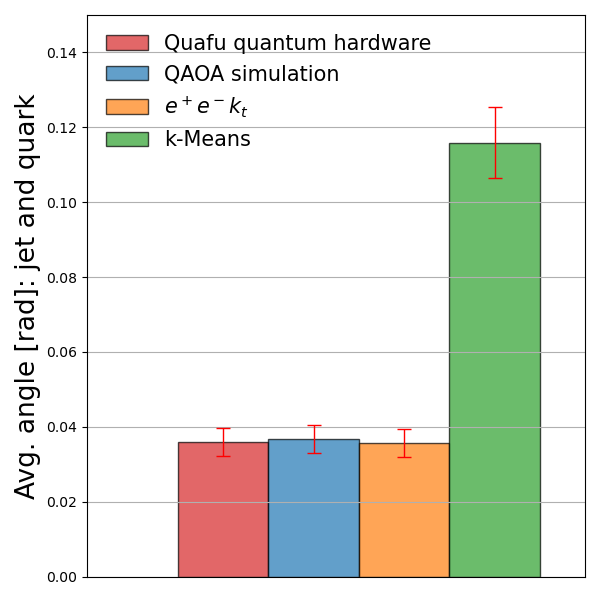}
    \caption{Figure taken from ~\cite{Zhu:2024own}:Jet clustering performance for 6-particle events on the \texttt{Baihua} quantum hardware is compared to that of quantum simulators, the classical $e^+e^-k_t$ algorithm, and the classical k-Means. \label{fig:quafu}}
\end{figure}

In each collision, the produced quarks and gluons immediately form collimated particle sprays known as jets. 
Jet clustering, which groups final-state particles into jets, is crucial as it retains the information of the originating quark or gluon and forms the basis for studying properties of the Higgs boson, as well as searching for new physics. 
Several studies~\cite{deLejarza:2022vhe,Delgado:2022snu,Pires:2020urc} have pursued potential quantum advantage in jet clustering, using quantum k-Means/$k_T$ algorithm or quantum annealing. 
By mapping collision events into graphs—with particles as nodes and their angular separations as edges, Ref~\cite{Zhu:2024own} presents a novel quantum realization of the jet clustering using the Quantum Approximate Optimization Algorithm (QAOA).
With up to 30 qubits used on a noiseless quantum simulator,
the performance of the QAOA method, based on the angle between the reconstructed jet and the corresponding quark—which ideally should be collimated—was found to be comparable to that of the classical $e^+e^-k_t$ algorithm and superior to that of the  classical k-Means, as shown in \fig{quafu}, 
The QAOA methods are further implemented on the \texttt{Baihua} quantum real machine on Quafu \cite{bib40} using 6 qubits. 
Mapping jets to graphs also allows jet clustering to be formulated as a quadratic unconstrained binary optimization (QUBO) problem, which quantum algorithms are expected to solve efficiently. 
A QUBO-based multijet reconstruction has recently been implemented using novel quantum annealing-inspired algorithms (QAIA)~\cite{Okawa:2024goh}.
These study highlights the feasibility of quantum computing to revolutionize jet clustering.

Track reconstruction (reconstruction of charge particle trajectories with hits in the inner detectors of collider experiments) is also of crucial importance.
It consumes significant computing resources, especially for HL-LHC and future hadron colliders. 
To tackle this challenge, various quantum algorithms, including QVE, QAOA and quantum graph neural networks (QGNN), are investigated for track reconstruction~\cite{Bapst:2019llh,Zlokapa:2019tkn,Tuysuz:2021oai, Okawa:2023doi, Crippa:2023ieq}.
For example recently, Ref~\cite{Okawa:2024eof} shows that the simulated bifurcation algorithm, a QAIA running on classical computers to solve QUBO problems, can be valuable for track reconstruction. 
Compared to the simulated annealing, the simulated bifurcation algorithm exhibits compatible or better reconstruction efficiency and purity, but the running time can be reduced by up to four orders of magnitude.

\textit{Quantum generative models ---}\label{sec:quantum-generative-models}
To derive high precision results, data analysis in HEP requires enormous amounts of simulated data. Monte Carlo methods are typically used in simulating the event generation and detector responses, which can be computationally demanding. This issue is particularly relevant to motivate recent studies of Generative Adversarial Networks (GAN), which are shown to exhibit remarkable speed-ups over Monte Carlo-based methods and increase the statistics of the produced Monte Carlo datasets \cite{Hashemi:2019fkn}. With the further increasing demands in the size of simulated datasets for the forthcoming colliders, quantum generative models might be promising approaches to further alleviate the strain on computational resources.

Advances have been seen recently in the development of quantum-assisted models for Monte Carlo event generation and for detector simulation. It is shown that the quantum generator architectures in the context of GAN for Monte Carlo event generation \cite{Bravo-Prieto:2021ehz} are able to learn the kinematic distributions of particles produced at LHC together with the benefits of higher training efficiency in terms of network depth. Complex correlations in the data can also be captured by quantum generative models such as quantum circuit-born machines (QCBM) that are difficult for classical models to learn, as studied in \cite{Delgado:2022aty}. Detector simulations with assistance from quantum algorithms are mainly focusing on simulating the calorimeter response, where reproducing the images of a reduced sizes of calorimeter responses has been achieved \cite{Chang:2021ufg}. Prototypes of quantum generative models with continuous variable quantum computing are being suggested with the necessity of further explorations to demonstrated its efficacy \cite{Chang:2021jne}. These algorithms on NISQ devices are being tested where optimizations are needed to reduce the effects of quantum noises \cite{Chang:2022dxc, Rehm_2024, baglio2024}.

\textit{Classification tasks --- }\label{sec:classification-tasks}
Accurate classification of collider events is crucial to separate rare signal from the immense background, and unlock discoveries. The massive datasets produced at future collider experiments contain billions of events with intricate, high-dimensional patterns, which will pose significant challenges for classical algorithms in processing time, resource demands, and recognizing particle correlations. QML holds the potential to handle such large, complex datasets more efficiently, and enable faster and more accurate classifications. 

Starting from the application of quantum annealing machine learning to optimize the classifiers for distinguishing Higgs events and background events in Ref. \cite{nature-qml}, exploring the feasibility and advantages of QML methods in the classification tasks has received increasing attentions. Modifications to the quantum annealing machine learning in classification tasks are studied,
reducing the performance gap to the classical counterpart at large training sample sizes \cite{QAML-Z}. Two different QML methods --Quantum variational classifier and support vector machine with a quantum kernel estimator (QSVM-Kernel method) -- have been applied to a flagship analysis (Higgs boson production in association with a top quark pair) at LHC \cite{Wu_2021, PRR-QSVM}, where comparable performance with classical machine learning methods are observed. These QML methods including hybrid quantum-classical graph convolutional network (QGCNN) have also been explored in jet classification \cite{Gianelle:2022unu}, particle identifications \cite{Chen:2021ouz, Yao:2024pij, Li:2023qsk, Yao:2024xlw}, measurements of physical processes at existing and future lepton colliders \cite{QML-CEPC, Ding:2024lfr}, where comparable discrimination power to other traditional machine learning models are usually found. A variational
quantum searching neighbor algorithm has also been proposed for identifying signals from new physics in the framework of effective field theories \cite{Yang:2024bqw}.
Despite the limitations of quantum computers, such as a limited number of qubits and high quantum error rates, the feasibility of quantum machine learning in classification tasks has begun to be established with these comprehensive studies.

\textit{Anomaly detection problem --- }\label{sec:anomaly-detection}
The absence of new physics discoveries at the LHC has motivated the exploration of model-independent data-analysis methods to efficiently identify anomalous events that could provide hints of new physics. Unsupervised machine learning techniques, also referred to as anomaly detection, are thus being widely studied \cite{Feickert:2021ajf}. Two neural network approaches—autoencoders, which are trained to learn only background distributions, and weakly supervised methods, which are trained to distinguish data from background-only samples—have become major research focuses in anomaly detection. The quantum counterpart of anomaly detection are also emerging along the advances of QML. Quantum anomaly detection based on autoencoder are explored to detect anomalous events at LHC \cite{Ngairangbam:2021yma, Wozniak:2023xbe, Duffy:2024zog, Bordoni:2023lad}, where better performance in the detection accuracy are mostly observed over their corresponding classical methods. Interestingly, it has also been demonstrated that such advantages can be attributed to the inherent properties of the quantum circuits, such as quantum entanglements \cite{Wozniak:2023xbe, Duffy:2024zog}. Different methods of embedding data into qubits are also being explored to further improve the detection accuracy as well as the representability of the data \cite{Araz:2024lsl}. Improvements have also been observed recently in anomaly detection using quantum similarity learning, surpassing its counterpart \cite{hammad2024}. Nevertheless, studies that apply quantum anomaly detection in the context of weakly supervised learning—specifically in the four lepton final state at the LHC—have yet to show clear advantages, which might be due to the specific nature of the problem, the limited size of the training dataset, or the particular quantum algorithms used \cite{Alvi:2022fkk}.

To recap, QML has been significantly tested for its applicability in data analysis within HEP. In certain scenarios, QML demonstrates improvements in training efficiency and data analysis accuracy over classical counterparts, particularly in object reconstruction and anomaly detection. Emerging studies aim to explore the relationship between quantum properties and these observed improvements, showing immense potential. Looking ahead, one of the research focuses in the HEP community will be testing the empirical claims of quantum advantages, with increased training sample sizes and the number of qubits used, the development of new QML algorithms tailored for HEP, and new quantum computing frameworks such as continuous variable or qudit systems. These efforts are likely to yield valuable and profound findings.

%% file: sections/quantum-pheno.tex
\section{Probing quantum nature at colliders}
\label{sec:pheno}
The research scope of HEP has expanded significantly at the high-energy frontier, thanks to systematic measurements and searches across thousands of sensitive phase spaces following the discovery of the Higgs boson in 2012. One intriguing research direction in HEP that has emerged is the study of the quantum nature (QN) of microscopic interactions at high-energy colliders \cite{Barr:2024djo}. This allows for an exploration of the fundamental properties of quantum mechanics in the unprecedented high-energy regime, offering a potential tool or even a new philosophical framework that could guide physicists in making future advancements in BSM searches. The study of quantum mechanics in HEP experiments focuses on fundamental properties such as quantum entanglement and the Bell inequality \cite{epr}, which are key features of quantum states and some of the most distinctive aspects of quantum mechanics.

On the other hand, null results of BSM physics have been seen in the numerous searches conducted at high-energy colliders such as the LHC. Searches so far have mostly relied on the measured cross-sections or differential kinematic distributions. A natural move for future search programs is to add further dimensions of sensitive information to enhance search sensitivity, such as introducing quantum effects as additional observables. This can involve quantum entanglement between initial and final states, spin correlations between particles, quantum interference effects, etc. In the following, we will give a brief recap of the definitions of quantum entanglement and the Bell inequality, and demonstrate how these can be measured at colliders. Examples from the latest measurements at the LHC will be shown. Then we will move on to discuss the practice of extending current measurements and searches with these new dimensions of quantum nature and further depict several preliminary ideas on how these research directions will evolve in the future.

\subsection*{Quantum entanglement and bell inequality}
In quantum mechanics, the general description of an $n$-level quantum state with its Hilbert space $\mathcal{H}_n$ is encoded in the density matrix, $\rho$. In this way, any physically accessible information of the system is obtained by computing
expectation values of observables: $\langle \mathcal{O}\rangle = \mathrm{Tr}\{\rho\mathcal{O}\}$.

\textit{Quantum entanglement ---}
For a quantum state, the subsystems Alice~($A$) and Bob~($B$) are entangled if $\rho$ cannot be written as a separable state of the form
\begin{equation}
\rho_{\rm } = \sum_i p_k \rho^{A}_{k} \otimes \rho^{B}_{k}, \label{eq:qe_rho_sep}
\end{equation}
where $p_k > 0$ are classical probabilities, with $\sum_k p_k = 1$, and $\rho^{A}_{k}$, $\rho^{B}_{k}$ are density matrices acting in the Alice and Bob Hilbert spaces. 
A general test for $\rho$ to determine whether it corresponds to a separable or an entangled state is not known. The most popular one is the Peres-Horodecki criterion~\cite{Peres:1996dw,Horodecki:1997vt},
which is a necessary condition and only sufficient for a bipartite system in two cases~\cite{Woronowicz:1976165}:  $dim \mathcal{H}_A = dim \mathcal{H}_B = 2$ (qubits) and
$dim \mathcal{H}_A = 2, dim \mathcal{H}_B = 3$ (and vice versa).

In practice, entanglement witnesses that can be easily computed are usually considered, and one widely used example is the concurrence $\mathcal{C}[\rho]$, with its value ranging from zero (for separable, unentangled states) to 1 (for maximally entangled states).  
For bipartite systems of a pair of qubits ($n = 2 \times 2$), $\mathcal{C}[\rho]$ is analytically
calculable~\cite{Bennett:1996gf,Wootters:1997id}. It can be found using the auxiliary $4 \times 4$ matrix $R=\rho \,  (\sigma_y \otimes \sigma_y) \, \rho^* \, (\sigma_y \otimes \sigma_y)$~\cite{Wootters:1997id}
with $\rho^*$ being the complex conjugate of $\rho$. Although non-Hermitian, $R$
possesses non-negative eigenvalues; denoting their square roots as $r_i$ with $i=1,2,3,4$,
and assuming $r_1$ to be the largest,
$\mathcal{C}[\rho]$ can be expressed as~\cite{Wootters:1997id}
\begin{equation}
\mathcal{C}[\rho] = \max \big( 0, r_1-r_2-r_3-r_4 \big)\ .
\label{concurrence}
\end{equation}

For bipartite systems of dimension larger than $2\times3$, $\mathcal{C}[\rho]$ is not analytically calculable. Fortunately, lower bounds on $\mathcal{C}[\rho]$ for a generic density matrix $\rho$ have been determined and,
if non-vanishing, unequivocally signal the presence of entanglement.
One computable bound among these is $\big(\mathcal{C}[\rho]\big)^{2} \geq \mathcal{C}_2[\rho]$~\cite{PhysRevLett.98.140505} where 
\begin{equation}
\mathcal{C}_2[\rho] = 2 \,\text{max}\, \Big( 0,\, \Tr[\rho^{2}] - \Tr[\rho_A^2],\, \Tr[\rho^{2}] - \Tr[\rho_B^2]  \Big) \ .
\label{eq:C_2}
\end{equation}
A non-vanishing $\mathcal{C}_2[\rho]$ indicates $\mathcal{C}[\rho]>0$, and therefore a non-zero entanglement content of $\rho$.

\textit{Bell inequality ---} 
For a Bell operator $\mathcal{B}$, its expectation value should satisfy the following Bell inequality 
\begin{equation}\label{eq:bell_obell}
\mathcal{I} = \langle \mathcal{B} \rangle = \mathrm{Tr}\{\rho\mathcal{B}\} \leq 2.
\end{equation}
for a local realist theory. For a pair of qubits, the Bell inequality can be written as the Clauser-Horne-Shimony-Holt (CHSH) form~\cite{Clauser:1969ny}
\begin{equation}
\label{eq:bell_chsh}
\mathcal{I}_2 =E(a, b)-E(a, b') + E(a', b) + E(a', b') \leq 2\,.
\end{equation}
This equation characterizes the results $a$, $b$ of experiments performed on systems A and B respectively,
where the primes indicate the results obtained from alternative versions of the experiment, such as by changing the axis of a spin measurement. 

For a pair of qutrits, the optimal~\cite{Masanes:2003jwa} Bell inequality is of the Collins-Gisin-Linden-Massar-Popescu (CGLMP) form~\cite{Collins:2002sun}.
To construct it, one again considers two observers $A$ and $B$, each having two measurement settings: $A_1$ and $A_2$ for $A$, and $B_1$ and $B_2$ for $B$, but with each experiment now having three possible outcomes.
One denotes by $P(A_i=B_j+k)$ the probability that the outcomes $A_i$ and $B_j$ differ by $k$ modulo $3$. 
Then the Bell inequality in the CGLMP form is written as
\begin{eqnarray}
\label{eq:bell_cglmp}
 \mathcal{I}_3 = && P(A_1=B_1) + P(B_1=A_2+1) + P(A_2=B_2)\notag\\ 
 &&+ P(B_2=A_1) - P(A_1=B_1-1) - P(B_1=A_2) \notag\\
 &&- P(A_2=B_2-1) - P(B_2=A_1-1)\leq 2.
 \end{eqnarray}

\subsection*{Quantum entanglement and Bell inequality at colliders}
The quantities used to test quantum entanglement and the violation of the Bell inequality can be derived from the spin density matrix $\rho$ of the system, which needs to be reconstructed for a given physical process, e.g., from observables of the final-state particles. This reconstruction also aligns with the long-established methods for studying helicity and polarization amplitudes at colliders. 

\textit{Qubits --- }The density matrix for a pair of spin-1/2 particles can then be expressed as
\begin{eqnarray}
\label{eq:rho-qubit}
\rho &=&\frac{1}{4}\Big[\mathbb{1}_2\otimes\mathbb{1}_2 + \sum_{i=1}^3 B^{+}_i (\sigma_i\otimes\mathbb{1}_2)\notag\\
&+&\sum_{i=1}^3 B^-_j (\mathbb{1}_2\otimes \sigma_j) + \sum_{i,j=1}^3 C_{ij} (\sigma_i\otimes\sigma_j) \Big]\ ,
\end{eqnarray}
where $\sigma_i$ are the Pauli matrices, $\mathbb{1}_2$ denotes the $2\times 2$ identity matrix and the indices $i, j= 1,2,3$ correspond to the three orthogonal directions in three-dimensional space. The coefficients
\begin{equation}
\label{rho-ab-coefficients}
B^+_i = {\rm Tr}[\rho\, (\sigma_i\otimes {\bf 1})] \quad \text{and} \quad
B_j^{-}={\rm Tr}[\rho\, ({\bf 1}\otimes\sigma_j)]\ ,
\end{equation}
represent the polarization
of each spin, and the matrix $C$
\begin{equation}
\label{rho-c-coefficients}
C_{ij} ={\rm Tr}[\rho\, (\sigma_i\otimes\sigma_j)]
\end{equation}
represents their correlations.
We then have:
\begin{eqnarray}
\mathcal{C}_2 &=& \frac{1}{2} \max \Big[ -1 +\sum_i (B_i^+)^2 -\sum_j (B_j^-)^2+ \sum_{i,j} C_{ij}^2, \notag\\ && -1 + \sum_j (B_j^-)^2 - \sum_i (B_i^+)^2 + \sum_{i,j} C_{ij}^2,\; 0\Big].
\end{eqnarray}
To test the Bell inequality, the two-spin state violates the CHSH inequality (\eq{bell_chsh}), if and only if the sum of the two greatest eigenvalues of $M\equiv CC^T$ is strictly larger than 1, according to the Horodecki condition.

\textit{Qutrits --- }The density matrix for a pair of qutrits can be expressed as
\begin{eqnarray}
\rho &=& \frac{1}{9}\left[\id\otimes
    \id\right]+
    \sum_{a=1}^8 f_a \left[T^a\otimes \mathbb{1}\right] \notag \\
    &+&\sum_{a=1}^8 g_a \left[\id\otimes T^a\right]
    +\sum_{a,b=1}^8 h_{ab}  \left[T^a\otimes T^b\right]\, ,
\label{eq:rho-qutrit-GM}
\end{eqnarray}
where the matrices $T^a$ are the standard Gell-Mann matrices. The coefficients $f_a$, $g_a$ and $h_{ab}$ are given by \begin{eqnarray}
f_a&=&\frac{1}{6}\,\Tr\left[\rho \left(T^a \otimes \mathbb{1}\right)\right]\, , ~~
g_a=\frac{1}{6}\,\Tr\left[\rho \left(\mathbb{1}\otimes T^a\right)\right]\notag \\
h_{ab}&=&\frac{1}{4}\,\Tr\left[\rho \left(T^a \otimes T^b\right)\right]\, .
\label{fgh}
\end{eqnarray}
Alternative ways to parametrize the $9\times 9$ density matrix can be found in \cite{Aguilar-Saavedra:2017zkn}. The observable $\mathcal{C}_2$ is given by:
\begin{eqnarray}
\mathcal{C}_2 &=& 2\max \Big[ -\frac{2}{9}-12 \sum_a f_{a}^{2} +6 \sum_a g_{a}^{2} + 4 \sum_{ab} h_{ab}^{2};\notag\\
  && -\frac{2}{9}-12  \sum_a g_{a}^{2} +6 \sum_a f_{a}^{2} + 4 \sum_{ab} h_{ab}^{2},\; 0 \Big]\,.
\label{eq:cmb-qutrit}
\end{eqnarray}
The Bell observable ${\cal I}_3$ to test the violation of Bell inequality can be written as
\begin{eqnarray}
\mathcal{I}_3 = && 4 \Big(h_{44} + h_{55} \Big) - \frac{4 \sqrt{3} }{3} \Big[ h_{61} + h_{66} \notag\\ &&+ h_{72} +  h_{77} +  h_{11} + h_{16} + h_{22} +  h_{27} \Big]\, . \label{eq:ii3}
\end{eqnarray}
\textit{Phenomenological studies and expected sensitivities ---}
The procedure to determine the density matrix of a quantum state by performing measurements over an ensemble of events is known as quantum tomography~\cite{White:1999sjn,James:2001klt,Thew:2002fom}. This method has been widely used in collider studies, and a complete formalism for the quantum tomography of the density matrix $\rho$ in general scattering processes has been developed in Ref.~\cite{Bernal:2023jba, Ashby-Pickering:2022umy}. The coefficients in the parametrization of $\rho$ can, in principle, be determined by averaging over the angular distribution of the final state in dedicated reference frames. Certain aspects unique to high-energy experiments need to be highlighted. In colliders, analyses allow each event to be reconstructed in different frames. Collecting the ensemble of events may not result in a quantum state, if the averaging is done in event-dependent frames. The object constructed in this manner has been labeled a ``fictitious state'' \cite{Afik:2022kwm}, which loses many characteristic properties of quantum states. Given the frame dependence on a per-event basis in collider experiments, there is an opportunity to determine optimal frame choices for maximizing spin correlations, entanglement, and Bell inequality violations \cite{Cheng:2023qmz, Cheng:2024btk}.
While the decays of particles are mostly considered for quantum tomography, a novel approach, referred to as the kinematic approach was recently proposed \cite{Cheng:2024rxi}.
This approach relies solely on the kinematics of particles to perform quantum tomography.
By leveraging the simplicity of scattering kinematics, this approach promises optimal statistical performance and offers the opportunity to establish Bell non-locality within existing data. Effects of higher-order corrections from QCD and EW in modifying the helicity structure in the production channel are also being scrutinized in \cite{Grossi:2024jae}. Building on these developments, along with those summarized in Table~\ref{tab:qe_pheno}, promising sensitivities can be achieved using existing or foreseen data that will be accumulated.
Below, we detail one example of measuring quantum entanglement at ATLAS and CMS.
\begin{table*}[ht!]
\centering
\begin{tabularx}{0.98\textwidth} { 
   >{\raggedright\arraybackslash}X |
   >{\centering\arraybackslash}X |
   >{\raggedleft\arraybackslash}X  }
 \hline\hline
 Process & Quantum entanglement~(QE) sensitivity & Violation of Bell inequality~(BI) \\ [0.5ex] 
 \hline
 \multicolumn{3}{c}{Qubits} \\
 \hline
 Top-quark pair production at the LHC\cite{Afik:2020onf,Fabbrichesi_2021,Severi_2022,Aguilar_Saavedra_2022,Aguilar-Saavedra:2023hss,Dong:2023xiw, Han:2023fci}& Expected to observe QE with LHC Run2 data~\cite{Afik:2020onf}& Expected to test the violation of BI at the 98\% CL with $\mathcal{L}=139$~fb$^{-1}$, setting aside certain systematic uncertainties~\cite{Fabbrichesi_2021}\\ \hline
 $\tau$-lepton pair production at the LHC~\cite{Fabbrichesi:2022ovb}, SuperKEKB~\cite{Ehataht:2023zzt} and FCC-ee \cite{Fabbrichesi:2024wcd} & \multicolumn{2}{p{0.5\textwidth}}{Expected to observe both QE and BI violation at $5~\sigma$ with exiting Belle~II data~\cite{Fabbrichesi:2022ovb}} \\ \hline
 Higgs boson decays in $\tau$-lepton pairs at CEPC~\cite{Ma:2023yvd}, ILC and FCC-ee~\cite{Altakach:2022ywa}& QE expected to be observed at $5~\sigma$ at both ILC and FCC-ee & BI violation expected to be measured at $3(1)~\sigma$ at FCC-ee~(CEPC)\\ \hline
 Higgs boson decays in di-photons~\cite{Fabbrichesi:2022ovb} & \multicolumn{2}{c}{Requires to measure the polarization of the two photons.} \\ \hline
 Hyperon pair production at BESIII~\cite{Wu:2024asu} & - & - \\ \hline
 B meson decays at LHCb~\cite{Gabrielli:2024kbz, Fabbrichesi:2023idl} & QE expected to be observed at $>5\sigma$ with present LHCb data & Expected to test BI violation at $>5\sigma$ with present LHCb data\\ \hline
 Charmonium decays at BESIII~\cite{Fabbrichesi:2024rec} & - & BI violated with a significance of $5\sigma$ or more in several decays such as $\eta_{c}, \, \chi_{c}^{0}, \, J/\psi \to \Lambda +\bar \Lambda \nonumber$ \\ \hline

 \multicolumn{3}{c}{Qutrits} \\ \hline
 Diboson~($WW, WZ, ZZ$) production at LHC~\cite{Ashby-Pickering:2022umy,Fabbrichesi:2023cev,Aoude:2023hxv, Grabarczyk:2024wnk} & Observation of QE might be possible & BI violation not expected\\ \hline
 $WW$ production at a 240~GeV $e^{+}e^{-}$ collider~\cite{Bi:2023uop}& - & Expected to test BI violation at $5~\sigma$ with $\mathcal{L}=180$~fb$^{-1}$
 (parton level) \\ \hline 
 $H\to WW^{*}$ at LHC~\cite{Barr:2021zcp,Ashby-Pickering:2022umy,Aguilar-Saavedra:2022mpg,Fabbrichesi:2023cev,Fabbri:2023ncz} & Expected to observe QE at $5~\sigma$ with LHC Run2 data~\cite{Aguilar-Saavedra:2022mpg}& Test the BI violation with a significance from $1\sim5\sigma$~\cite{Barr:2021zcp}\\ \hline
 $H\to ZZ^{*}$ at LHC~\cite{Aguilar-Saavedra:2022wam,Ashby-Pickering:2022umy,Fabbrichesi:2023cev,Bernal:2023ruk} & Expected to observe QE at $3~\sigma$ with $\mathcal{L}=300$~fb$^{-1}$~\cite{Aguilar-Saavedra:2022wam} & Expected to test BI violation at $4.5~\sigma$ with HL-LHC\\ \hline
 Vector-boson scattering $V_1^{'}V_2^{'}\to V_1 V_2$~\cite{Morales:2023gow} & - & - \\
 \hline
 \hline
\end{tabularx}
\caption{Representative theoretical and phenomenological studies on quantum entanglement and violation of Bell inequality at colliders. The `-' symbol indicates no explicit expected sensitivities are given.}
\label{tab:qe_pheno}
\end{table*}
\\
\\
The ATLAS Collaboration~\cite{ATLAS:2023fsd} has recently performed the first measurement of quantum entanglement in $t\bar{t}$ pair events produced at the LHC, 
based on a proton--proton collision dataset with a center-of-mass energy of $\sqrt{s}=13$~TeV and an integrated luminosity of 140~fb$^{-1}$. 
Spin entanglement is detected from the measurement of a single observable $D$ related to the concurrence through the relation $\mathcal{C}[\rho]=\max [-1-3D,\,0]/2$~\cite{Afik:2020onf}, which leads to the entanglement criterion: $D < -1/3$. The analysis selects fully leptonic top pair events with one electron and one muon of opposite signs. $D$ can be extracted from the differential cross section
\be 
\frac{1}{\sigma} \frac{\mathrm{d} \sigma}{\mathrm{d} \cos \phi} = \frac{1}{2} \Big( 1 - D \cos \phi \Big)\, ,
\ee
with $\phi$ being the angle between the two leptons in the rest frame of the decaying $t\bar{t}$ pair. The measurement is performed in a narrow interval around the $t\bar{t}$ threshold $340\,{\rm GeV} < m_{t\bar{t}}< 380$\,GeV, where the production predominantly arises from the spin-singlet state, which is maximally entanglement \cite{Han:2023fci}.

\begin{figure}[htbp]
    \centering
    \includegraphics[width=0.9\linewidth]{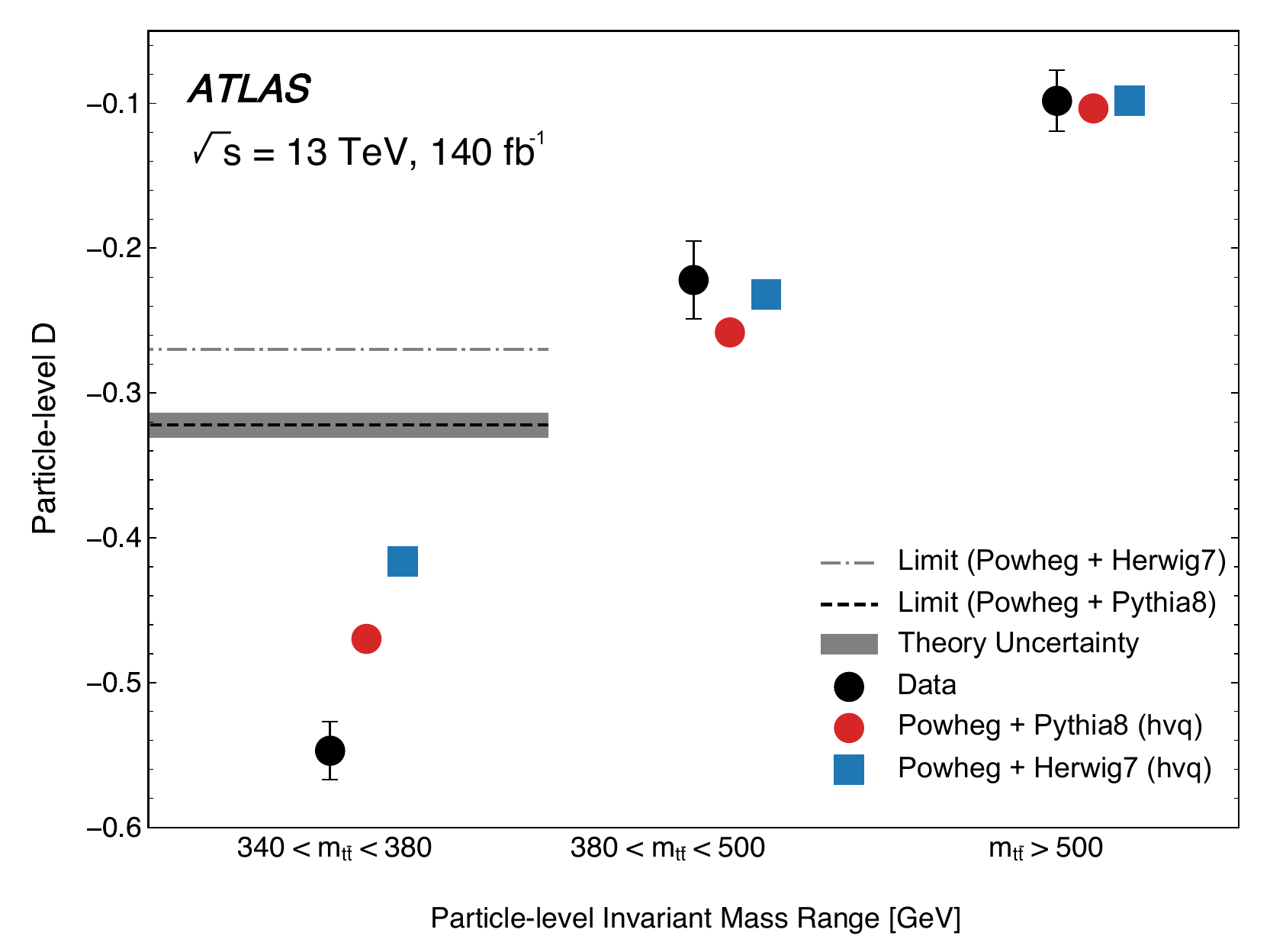}
    \caption{The observed values of the entanglement marker $D$ compared with various MC models. The entanglement limit shown is a conversion from its parton-level value of $D=-1/3$ to the corresponding value at particle level, and the band represents the uncertainties.}
    \label{fig:ATLAS_ttbar}
\end{figure}

After calibrating for detector acceptance and efficiency, they measured~\cite{ATLAS:2023fsd}
\be
D = -0.547 \pm 0.002\; [{\rm stat.}] \pm 0.021\; [{\rm syst.}]\, .
\ee
The observed and expected significances with respect to the entanglement limit ($D < -1/3$) are well beyond $5\sigma$, 
independent of the Monte Carlo models that could shift the parton-level entanglement limit to account for the fiducial phase space of the measurement, as shown in Figure~\ref{fig:ATLAS_ttbar}.
This result provides the first experimental observation of entanglement between a pair of top quarks. The CMS Collaboration~\cite{CMS:2024pts} has reported a similar measurement, yielding an observed~(expected) significance of $5.1~(4.7)~\sigma$ for the witness of quantum entanglement.

\subsection*{Quantum dimension for measurements and searches}
Quantum dimension collectively refers to the effects related to the QN of concerned physics processes, such as interference effects, spin, and polarization of final-state particles. The use of quantum dimension has historically been a key means of achieving theoretical and experimental discoveries in HEP. This was particularly successful in the early path to establishing the SM, with numerous examples such as the study of CP violation, determination of the CKM matrix, and hadron spectroscopy. At high-energy colliders, the opening of new channels and phase spaces has stimulated numerous research efforts into the unknown. The first wave of studies scrutinized the physics processes based on cross-sections and basic kinematic distributions, while future efforts will examine these processes with the largest possible datasets and a much sharper focus. Physics insights can naturally be gained by exploring the quantum dimension.

For example, to measure the couplings of the Higgs boson accurately, one must establish a full relationship between observables (such as cross-sections of different Higgs-related processes) and all coupling constants, where interference terms naturally appear and play important roles. This fact is clearly demonstrated in auxiliary Table 5 of Ref.~\cite{ATLAS:2022vkf}, where the interference contributions are the only means to determine the relative signs between coupling constants. These signs are key intrinsic properties that cannot be probed otherwise. Another example lies in the determination of the natural width of the Higgs boson, which is a fundamental parameter of vital importance but has a tiny SM value of $\unit[4.7]{MeV}$, well below the detector resolution for measuring the on-shell Higgs boson peak. In a recent measurement from the ATLAS experiment, the clean four-lepton final state originating from $ZZ$ diboson production is used to search for off-shell Higgs contributions, where the interference between the off-shell $H \rightarrow ZZ$ and the box-diagram $gg \rightarrow ZZ$ processes must be carefully evaluated and utilized for a delicate and sensitive measurement of off-shell Higgs production. This off-shell measurement is then combined with the on-shell measurement to achieve leading precision, with a width determination of about $\unit[3]{MeV}$~\cite{ATLAS:2023dnm}.

The second topic is the measurement of final-state particle polarization at colliders. Measuring boson polarization has a long history since the LEP and Tevatron experiments, where the focus was on the abundantly produced single vector boson process ($V = W, Z$) or a vector boson in association with jets ($V +$ jets). Polarization measurements in these channels significantly improve the understanding of electroweak and QCD theories, and in particular, improve the precision in determining fundamental parameters such as the weak mixing angle, as well as constraining parton distribution functions and higher-order QCD calculations. The production of two or more bosons (multi-bosons) is key to probing the gauge structure of the SM and searching for new physics, but this is not well measured due to insufficient statistics and limited collision energy. In the LHC era, the abundant high-energy collision data provides a unique opportunity for substantial progress in this area, and in fact, adding the polarization dimension to measurements helps to sharpen the physics sensitivity to the next level~\cite{Fabbrichesi:2023jep}. This can be easily understood from various physics discussions, for example: heavy resonances in many new physics models preferentially decay into two longitudinally polarized vector bosons ($V_{L}V_{L}$); the longitudinally polarized vector boson is directly related to the Goldstone boson equivalence theorem, which exhibits unique features enabling the probe of the relevant physics in greater depth.

Measuring the polarization of multi-bosons is not straightforward, as one needs to work with limited amounts of data and separate different states that are kinetically similar, in addition to differentiating between signal and background processes. Thanks to great improvements in the precision measurement of multi-boson processes, the polarization measurements start to sail, utilizing sensitive definitions of angular variables as well as the aid of machine learning techniques. For example, recent measurements of $Z_{L}Z_{L}$ and $W_{L}Z_{L}$ from the ATLAS experiment~\cite{ATLAS:2023zrv, ATLAS:2024qbd}. All polarization studies at present focus on the observation of different states, and it would be interesting to foresee that, in the future, decisive improvements to the HEP physics program could be achieved by adding such a key dimension.

%% file: sections/conclusions.tex
\section{Conclusions and outlook}
\label{sec:conclusion}

Quantum technology has increasingly permeated nearly every aspect of HEP research. Quantum sensing, as high-sensitivity detectors, has been widely considered in the exploration of hidden sectors and new phenomena such as dark matter searches and high-frequency gravitational wave detection. With its quantum properties, quantum computing, as a new computing paradigm, is being investigated for studying real-time, non-perturbative dynamics in the early universe and at colliders. The integration of quantum computing with machine learning is also being actively explored for applications in HEP data analysis, such as Monte Carlo event generation and object reconstruction. Meanwhile, testing fundamental properties of quantum mechanics at colliders is an emerging area, including tests of quantum entanglement and violations of the Bell inequality, with the entanglement of a two-qubit system at $\mathcal{O}(\unit[100]{GeV})$ energy scale recently observed at LHC. These phenomena also provide valuable observables in the quantum domain to search for new physics.

While significant progress has been made, further efforts are required in several areas. In sensing, advancements in reducing quantum noise are needed to enhance detection sensitivity. In computing, the development of theoretical frameworks and algorithms tailored to specific physics problems remains a priority. Additionally, coordinated efforts between the HEP and quantum hardware communities are essential to accelerate the implementation of quantum devices for computing HEP processes. In QML, improvements over their classical counterparts for anomaly detection and object reconstruction in HEP data analysis are starting to emerge; however, extensive research is still needed to identify the physics scenarios where QML can provide a clear advantage. While a plethora of phenomenological studies have been proposed to test quantum properties at colliders, the experimental program has just started, and more results are expected to be forthcoming. These measurements of quantum properties could also strengthen the search for new physics in the quantum domain, which remains to be explored. The integration of resources and collaborative efforts will be essential to fully leverage quantum technologies for HEP, ultimately driving deeper insights and enhancing our understanding of the fundamental laws of nature.